\documentclass[aps,prr,twocolumn,superscriptaddress,floatfix,notitlepage]{revtex4-2}

\usepackage{graphicx}
\usepackage{bbm}
\usepackage{float} 
\usepackage{color}
\usepackage[usenames,dvipsnames]{xcolor}
\usepackage{amsmath}
\usepackage{amssymb} 
\usepackage{amstext}
\usepackage{latexsym}
\usepackage[colorlinks=true,citecolor=NavyBlue,linkcolor=RubineRed,urlcolor=NavyBlue]{hyperref}

\newcommand{\figpanel}[2]{\hyperref[#1]{\ref{#1}#2}}

\usepackage{bm}

\begin{document}
\preprint{APS/123-QED}

\title{Annealing Dynamics of Regular Rotor Networks: Universality and Its Breakdown}

\author{Andr\'as Grabarits\href{https://orcid.org/0000-0002-0633-7195}{\includegraphics[scale=0.05]{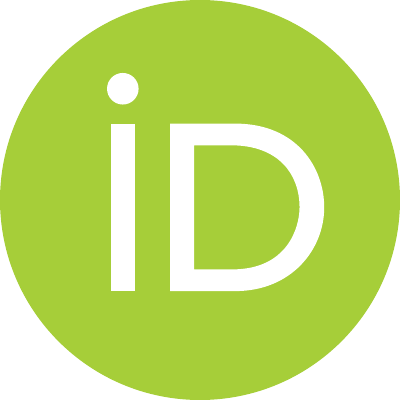}}}
\affiliation{Department  of  Physics  and  Materials  Science,  University  of  Luxembourg,  L-1511  Luxembourg, Luxembourg}

\author{Gaetano Sammartino\href{https://orcid.org/0000-0002-2681-9132}{\includegraphics[scale=0.05]{orcidid.pdf}}}
\affiliation{Department  of  Physics  and  Materials  Science,  University  of  Luxembourg,  L-1511  Luxembourg, Luxembourg}

\author{Adolfo del Campo\href{https://orcid.org/0000-0003-2219-2851}{\includegraphics[scale=0.05]{orcidid.pdf}}}
\affiliation{Department  of  Physics  and  Materials  Science,  University  of  Luxembourg,  L-1511  Luxembourg, Luxembourg}
\affiliation{Donostia International Physics Center,  E-20018 San Sebasti\'an, Spain}
\date{\today}

\begin{abstract}
The spin-vector Monte Carlo model is widely used as a benchmark for the classicality of quantum annealers but severely restricts the time evolution. The spin-vector Langevin (SVL) model has been proposed and tested as an alternative,  closely reproducing the real-time dynamics of physical quantum annealers such as D-Wave machines in the dissipative regime.
We investigate the SVL annealing dynamics of classical $\mathrm O(2)$ rotors on regular graphs, identifying universal features in the nonequilibrium dynamics when changing the range of interactions and the topology of the graph.
Regular graphs with low connectance or edge density exhibit universal scaling dynamics consistent with the Kibble-Zurek mechanism, which leads to a power-law dependence of the density of defects and the residual energy as a function of the annealing time. As the interaction range is increased, the power-law scaling is suppressed, and an exponential scaling with the annealing time sets in. Our results establish a universal breakdown of the Kibble-Zurek mechanism in classical systems characterized by long-range interactions, in sharp contrast with previous findings in the quantum domain. 
\end{abstract}

\maketitle

\section{\label{sec:intro}Introduction}
Quantum Annealing (QA) is a heuristic algorithm that exploits quantum effects to solve optimization problems \cite{Kadowaki98,Albash18,Hauke20}. QA relies on the gradual transformation of a quantum system from an initial and easily preparable state to a target state representing the solution of the optimization problem. Specifically, the solution is encoded as the ground state of the problem Hamiltonian. 
Due to the presence of noise and decoherence, current quantum annealing devices are best described as programmable many-body open quantum systems \cite{Boixo2013,Boixo2016,Mishra2018,Bando20}. A unitary time evolution provides an accurate description for short annealing times \cite{King22,King24} and shallow circuits in a digital implementation. These devices offer an exciting platform to explore non-equilibrium physics \cite{King24}. 
One central paradigm in this context is the Kibble-Zurek (KZ) mechanism, describing the dynamics of classical continuous and quantum phase transitions. It predicts that when a system is driven across a critical point in a time scale $\tau_Q$, adiabaticity breaks down, leading to the formation of topological defects at a density which varies as a universal power law with the quench time \cite{Kibble_1976, Kibble_1980, Zurek_1985}. In many-body spin systems of relevance to QA, defects generally represent errors in the optimization process that limit the preparation of low-energy states. 
The KZ mechanism has been used as a benchmark in quantum simulators \cite{Dziarmaga10,del_Campo_2014,Cui16,Keesling19,Cui20} and annealing devices \cite{Gardas18,Weinberg20,Bando20,Subires22,King22,King24}. In this context, efforts to identify and establish quantum signatures of the dynamics have focused on ruling out the behavior of classical models embedded in a thermal bath \cite{Kadowaki98,Smolin13,Wang13,Shin14,Albash15,AlbashLidar15,Albash21,Bando21,Subires22,Rajak23}. In addition, the KZ mechanism holds in classical spin systems with time-continuous simulated annealing and Glauber dynamics \cite{Krapivsky10,Jeong20,Mayo21,Fan23}, as well as with Monte Carlo updates \cite{Suzuki11,Bando20}. 

The conventional derivation of the KZ scaling, applicable to classical systems, relies on spatial locality \cite{Zurek96a,del_Campo_2014, DKZ13,delcampo22}, and yet, the KZ mechanism has been shown to hold in quantum systems with long-range interactions decaying as a power law. This is the case for nonintegrable systems, such as the transverse-field Ising model with power-law interactions \cite{Jaschke17,Sadhukhan20,Li2022probing} and related systems realizable with Rydberg atom quantum simulators  \cite{Keesling19,Chepiga21}, as well as in systems with a free-fermionic structure,  such as the long-range Kitaev model \cite{Dutta17}.  By contrast, fully connected systems exhibit a power-law scaling with the quench time, which is generally incompatible with KZ mechanism and better explained in terms of Landau-Zener crossings and quasi-adiabatic approximations \cite{Caneva08,Acevedo14,Defenu18,Gherardini23}.

The coupling of a quantum critical system to a bath need not destroy the KZ behavior and can preserve it by simply renormalizing the equilibrium critical exponents \cite{Silvi16,Bando20}.
However, environmental effects can give rise to
anti-Kibble-Zurek behavior, whereby the density of kinks increases with the annealing time, as first predicted in short-range systems \cite{Dutta16} and observed in several D-Wave annealing devices \cite{Gardas18,Weinberg20,King22}.
The anti-Kibble-Zurek scaling has been further established in the long-range Kitaev model  \cite{Singh21,king2022longrange,Singh23} and fully connected systems \cite{Puebla20}. 

In the classical domain, early studies in ion trap systems showed that power-law Coulomb interactions preserve the KZ scaling expected in short-range systems \cite{delCampo10,DeChiara10,Nigmatullin16}, with deviations reported in experiments in small samples \cite{Ulm13,Pyka13}. Likewise, the critical dynamics of the dipolar spin ice exhibits KZ scaling \cite{Hamp15,Fan23}. Recent experiments in systems with dipolar interactions reported the universal KZ  scaling when excitations are topological defects and found an enhanced suppression with the driving time in the case of nontopological excitations \cite{Du23}. Deviations from the KZ scaling are expected in the presence of coarsening dynamics \cite{Biroli2010,delcampo2024coarsening}, as has been described in systems with power-law interactions when crossing equilibrium  \cite{Libal20,Samajdar24} and nonequilibrium \cite{Reichhardt22,Reichhardt23} phase transitions.  

Power-law interactions decaying with the distance are natural in physical systems and can be studied in a variety of simulators using ultracold gases and Rydberg atoms \cite{Saffman10,Chomaz23}. Complex couplings involving nonlocal interactions arise in computational problems, statistical physics, and the study of networks.  Recent experimental progress has demonstrated the realization of programmable interactions, giving rise to a synthetic emergent geometry different from the physical one \cite{Periwal21}.

In spite of this progress, the validity of the KZ mechanism as a function of the interaction range in classical systems remains to be studied and is the focus of our work. We consider the annealing dynamics of a regular rotor network as a function of the connectance or edge density, i.e., the ratio between the number of edges and the maximal possible value associated with a fully connected graph. While a universal KZ scaling is observed in networks with small edge density, our results establish the breakdown of the KZ scaling as the connectance is increased, giving rise to the onset of an exponential suppression of the density of excitations with the annealing time. 

The theoretical framework for the classical Langevin annealing dynamics of $O(2)$ rotor networks is introduced in Sec. \ref{sec:SVL_and_Graph}, which also discusses the equilibrium relaxation time and magnetization.  This sets the ground for the study of KZ scaling
in graphs with low connectance in Sec. \ref{sec:KZM} and its breakdown at high edge density in Sec. \ref{SecKZMbreaks}. Universal signatures beyond the KZ mechanism characterizing fluctuations of the defect and energy densities are discussed in Sec. \ref{sec:beyondKZM}. Section \ref{sec:conclusions} closes with a summary and outlook.

%
%
\section{\label{sec:SVL_and_Graph} Time-evolution and equilibirum properties}
This section summarizes the details of the classical stochastic dynamics of rotor networks and the associated equilibrium statistical properties.

\subsection{\label{sec:SVL}The Spin Vector Langevin model}

The QA algorithm relies on the dynamics of a system generated by a Hamiltonian of the form
\begin{equation}
    H(t) = A(t)H_0+B(t)H_P,
\end{equation}
where $A(t)$ and $B(t)$ define the annealing protocol interpolating between a trivial Hamiltonian $H_0$ and the problem Hamiltonian $H_P$. These functions are chosen to satisfy the boundary conditions  $A(0)=B(\tau_Q)=1$ and $B(0)=A(\tau_Q)=0$, where $\tau_Q$ is the duration of the process.
Here, $H_0$ and $H_P$ are usually Ising type Hamiltonians of the form
\begin{align}
    &H_0 = -\sum_{i \in V}\sigma^x_i,\\
    &H_P = -\sum_{(i,j)\in E}J_{ij}\sigma_i^z\sigma_j^z -\sum_{i\in V}h_i\sigma_i^z,
\end{align}
with $\sigma^z_i$ and $\sigma^x_i$ denoting the $z$ and $x$ Pauli matrices on the $i$-th site.
The QA algorithm aims to find the solution to combinatorial optimization problems, which are encoded in the ground state of $H_P$. The classical hardness of these problems is characterized by both the underlying graph topology, $\mathcal{G}(V,E)$, and the spin-spin interactions $J_{ij}$~\cite{Lucas_Np_hard_review}. Here, $V$ is the set of vertices, and $E$ denotes the set of edges accounting for the presence of interactions among the vertices. In the course of the QA algorithm, the system starts from the easy-to-prepare ground state of $H_0$, and under slow adiabatic dynamics, it ends up in the final ground state. However, in quantum systems, the classical hardness of the given optimization problem leads to small energy gaps at quantum critical points. As a result, this leads to time scales similar to those required by classical algorithms for successful QA processes.

Even though the understanding of the crossing of such critical points in finite-range models is still in its infancy, valuable insights are offered by the stochastic Langevin dynamics of classical systems~\cite{delCampo10,DeChiara10,Gardas18, Bando20,Subires22, King22}.
To this end, we explore the finite-range extension of the ring topology of classical rotors, where the interactions are described by circulant graphs.
The corresponding spins are represented by classical spin vectors with trigonometric functions replacing the Pauli operators as $\sigma_i^z \rightarrow \sin\theta_i$ and $\sigma_i^x \rightarrow \cos\theta_i$. The angles $\theta_i$ describe classical planar $O(2)$ rotors characterized by the $N$-dimensional vector $\bm\theta = (\theta_1,\theta_2,\ldots,\theta_N)$. 
Note that this mapping has been justified to reproduce most of the low-energy physics in the corresponding quantum spin systems with tunable power-law interactions, and common benchmarks of classicality rely on it
\cite{Smolin13,Wang13,Shin14,Albash15,Albash21,Bando20,Bando21,Subires22,Rajak23}.
The corresponding finite-range interacting Hamiltonian is given by
\begin{equation}\label{eq:Hamiltonian}
	H(\bm{\theta},t) = -\frac{J(t)}{2}\sum_{i,j=1}^N A_{ij}\sin\theta_i\sin\theta_j -h(t)\sum_{i=1}^N\cos\theta_i\,.
\end{equation}
The annealing schedule specifies the time dependence of the Hamiltonian \eqref{eq:Hamiltonian}, and in real QA machines, it is determined by the balance of engineering optimization and the physical limitations of the hardware.
We choose a linear protocol 
\begin{equation}\label{eq:schedule}
    \quad J(t)=J_0\frac{t}{\tau_Q}\quad\text{and}\quad h(t) = h_0\left(1-\frac{t}{\tau_Q}\right) ,t\in[0,\tau_Q]
\end{equation}
where $J_0$ and $h_0$ are arbitrary constants fixing the energy scale of $H_P$ and $H_0$, respectively.
Here $A_{ij}$ denotes the elements of the adjacency matrix describing the circulant graphs $\mathcal G(V, E)=C_{i_N}(r)$. Here, $N$ denotes the number of vertices of the graph, i.e., the number of rotors, $\sum_{i<j}\,A_{ij}=\lvert V\rvert=N$. The integer $r$
sets the interaction range between spins,
\begin{equation}\label{eq: A_ij_def}
    A_{ij} = 
    \begin{cases}
    1 &\text{if} \;\, |i-j| \le r,\\
    0 &\text{otherwise.}
    \end{cases}
\end{equation}
In Eq.~\eqref{eq:Hamiltonian}, the first term takes the role of the $H_P$ problem Hamiltonian, while the second is that of the $H_0$ transverse field Hamiltonian.
Circulant graphs provide the simplest generalization of the ring topology by extending the interaction range to the $r$ nearest neighbors. From the point of view of graph theory, the distance between two vertices corresponds to the number of connecting edges in the shortest path linking them. This shrinks the distances between vertices, leading to a non-trivial geometrical structure. However, a one-dimensional interpretation provides an accurate description of the non-equilibrium dynamics near the critical point.

\begin{figure}
\centering
\includegraphics[width=0.41\textwidth]{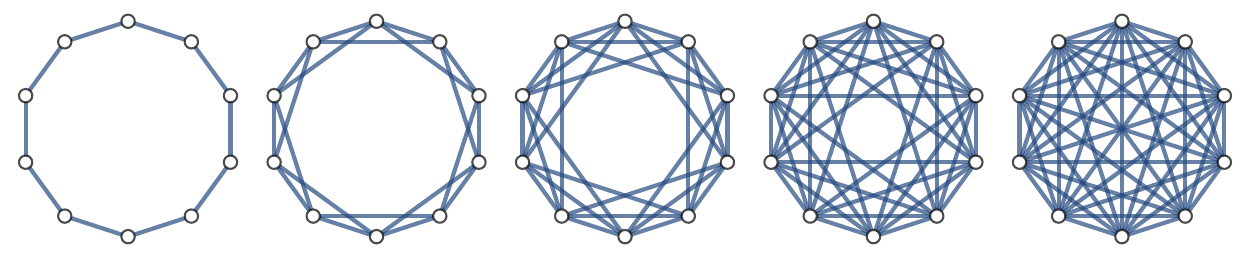}
\caption{\label{fig:graphs} Schematic representation of $O(2)$ networks associated with circulant graphs of order $N=10$ and increasing connectance from left to right, interpolating between the cycle graph $C_N$ and the complete graph $K_N$.}
\end{figure}

Due to the homogeneity of the system, these graphs are $2r$-regular. For simplicity,  we will refer to $r$ as the regularity in the following. 
To characterize these graphs independently of the system size, the graph density, or connectance, is introduced: $c = \frac{2r}{N-1}$. The connectance interpolates between the empty graph ($c = 0$) and the complete graph $K_N$ with $c = 1$ \cite{Newman_2018}, as depicted in Fig.~\ref{fig:graphs}.

The corresponding dynamics can be implemented in different ways.
The Spin-Vector Monte Carlo (SVMC) model is a convenient method to implement classical stochastic dynamics, capturing time evolution via discrete Monte Carlo steps \cite{Wang13,Shin14}. While it remains a common benchmark for annealing devices \cite{Albash15,Albash21,Bando20,Bando21,Rajak23}, it is not suited to describe the real-time dynamics unambiguously. Its evolution is highly restricted, being parameterized solely by temperature and lacking the notion of dissipation. The Spin Vector Langevin (SVL) model replaces the Monte Carlo steps with continuous-time Langevin dynamics that is flexible enough to account for dissipative effects by including a friction coefficient subject to the fluctuation-dissipation theorem \cite{Subires22}. It also provides a remarkable improvement over the SVMC method giving access to the real-time evolution of the system. 

The SVL evolution is encoded in the set of stochastic coupled differential equations
\begin{equation}\label{eq:Langevin}
\begin{split}
    &m_i\ddot{\theta_i}+\gamma_i\dot{\theta_i} +\frac{\partial H(\bm\theta,t)}{\partial \theta_i}+\xi_i(t)=0\,,\;\; i=1,2,\ldots,N\,\\
    & \frac{\partial H(\bm\theta,t)}{\partial\theta_i}=-J(t)\sum_{j=1}^NA_{ij}\cos\theta_i\sin\theta_j +h(t)\sin\theta_i,
    \end{split}
\end{equation}
where $\xi_i (t)$ represents independent and identical Wiener processes with zero mean, describing white noise on the $i$-th vertex. This noise generates classical fluctuations modeling how the external thermal bath affects the dynamics ~\cite{Bando20,Smolin13,Wang13}.
The effective mass of the rotors $m_i$ weights the inertia term, and $\gamma_i$ denotes the local damping constant. In the following, a homogeneous mass and damping are chosen, setting $m_i=m$ and $\gamma_i=\gamma$. Then, the noise autocorrelation is $\langle\xi_i(t)\xi_j(t')\rangle= 2m\gamma k_B T \delta_{ij}\delta(t-t')$, as required by the fluctuation-dissipation theorem \cite{Kubo66} for a thermal bath at temperature $T$. In this work, natural units are adopted for simplicity, setting $k_B=1,\,m=1$, and for the temperature, $T=0.001J_0$.
 As for the numerical implementation, the Langevin equations are integrated with the help of the multi-dimensional explicit order $2.0$ weak scheme of Ref.~\cite{Kloeden11,Subires22}. 
%
%

%
%
%
\subsection{\label{sec:schedule}Equilibrium critical exponents and order parameter}

As in the ring topology with $r=1$, regular rotor networks undergo a paramagnetic-ferromagnetic-like phase transition at values of $J(t)$ and $h(t)$ for which the two competing terms in the Hamiltonian are of the same order of magnitude for arbitrary $N$ and $r$.  The system starts from the paramagnetic phase, where the angles are approximately paramagnetically aligned $\theta_i\ll1$. Similarly, in the regular network, the Hamiltonian can be approximated by the quadratic form $H(\bm\theta)\approx-\frac{J(t)}{2}\sum_{i,j=1}^N A_{ij}\theta_i\theta_j-h(t)N+\frac{h(t)}{2}\sum_{i=1}^N\,\theta^2_i$. 
 The corresponding Langevin equations read
\begin{equation}\label{eq:linLangevin}
    m\ddot{\theta_i}+\gamma\dot{\theta_i} +h(t)\theta_i - J(t)\sum_{\substack{j=-r \\ j\neq 0}} ^r\theta_{i+j}+\xi_i(t)=0\,,\; \forall i.
\end{equation}
From Eq. \eqref{eq:linLangevin}, both the critical point and the critical exponent of the relaxation time can be determined in the different damping regimes \cite{Laguna98, delCampo10, DeChiara10, Subires22}. In the overdamped regime,  using the approximation  $\gamma\dot\theta\gg\ddot\theta$, one finds
\begin{equation}\label{eq:overdamped}
\tau\sim\left|\theta/\dot{\theta}\right|\sim\frac{\gamma}{h(t)-2rJ(t)}\,,
\end{equation}
setting the critical point at $h(t)=2rJ(t)$. To make the corresponding critical time independent of the connectance, $J_0=1$ and $h_0=2r$ are chosen, yielding $t_c=\tau_Q/2$.
In the underdamped case, the inertial term $\ddot\theta$ dominates and the equilibrium relaxation time reads
\begin{equation}\label{eq:underdamped}
\tau\simeq\left|\theta/\ddot{\theta}\right|^{1/2}\simeq\left|\frac{m}{h(t)-2rJ(t)}\right|^{1/2}.
\end{equation}
These equations imply that the regularity does not affect the relaxation time beyond shifting the critical point.
The corresponding critical exponents obey $z\nu=1$ in the overdamped regime and $z\nu=1/2$ in the underdamped regime.
In both cases, the natural choice for the control parameter is $\epsilon(t) = h(t)-2rJ(t)$, so that the transition takes place at $\epsilon(t_c) = 0$.
Shifting the beginning of the annealing schedule by $\tau_Q/2$ leads to $t_c \equiv 0$, convenient for further analysis. This way, the annealing schedule extends over the interval of $[-\tau_Q/2, \tau_Q/2]$, with the linearized control parameter given by $\epsilon(t)=2r\,t/\tau_Q$ around the critical point. Consistently, near $t_c$, $H_P$ and $H_0$ vary over the same energy scale, independently of $N$ and $r$. The quadratic part scales as $\sum_{i,j}A_{ij}\sin\theta_i\sin\theta_j \sim Nr$, while $H_0$ acquires the same order of magnitude thanks to the additional scaling of $h_0$ in $r$, $h_0\sum_{i}\cos\theta_i \sim Nr$.

\begin{figure}
\includegraphics[width=\linewidth,trim={2cm 5cm 4cm 0},clip]{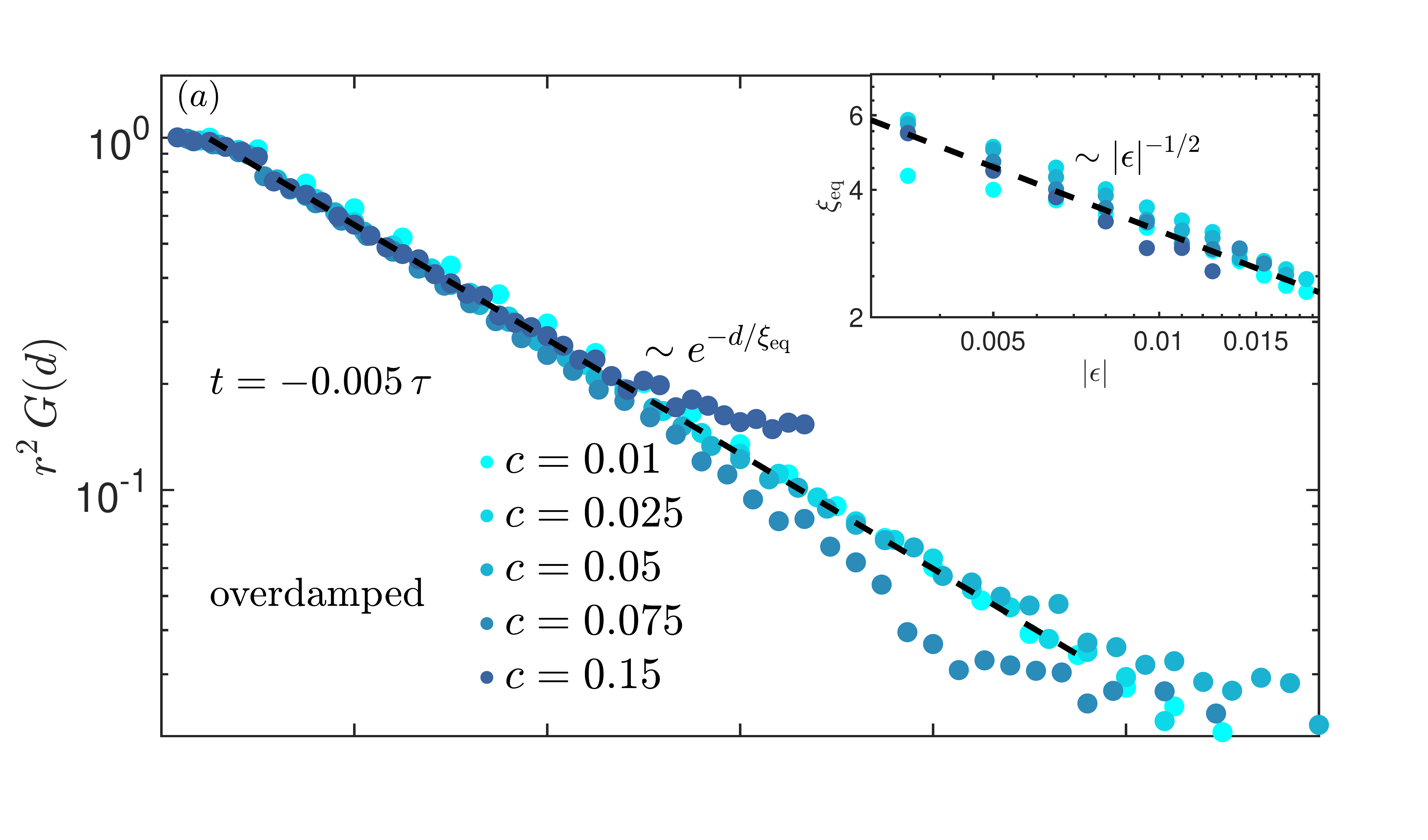}
\includegraphics[width=\linewidth,trim={1cm 0cm 4cm 4.2cm},clip]{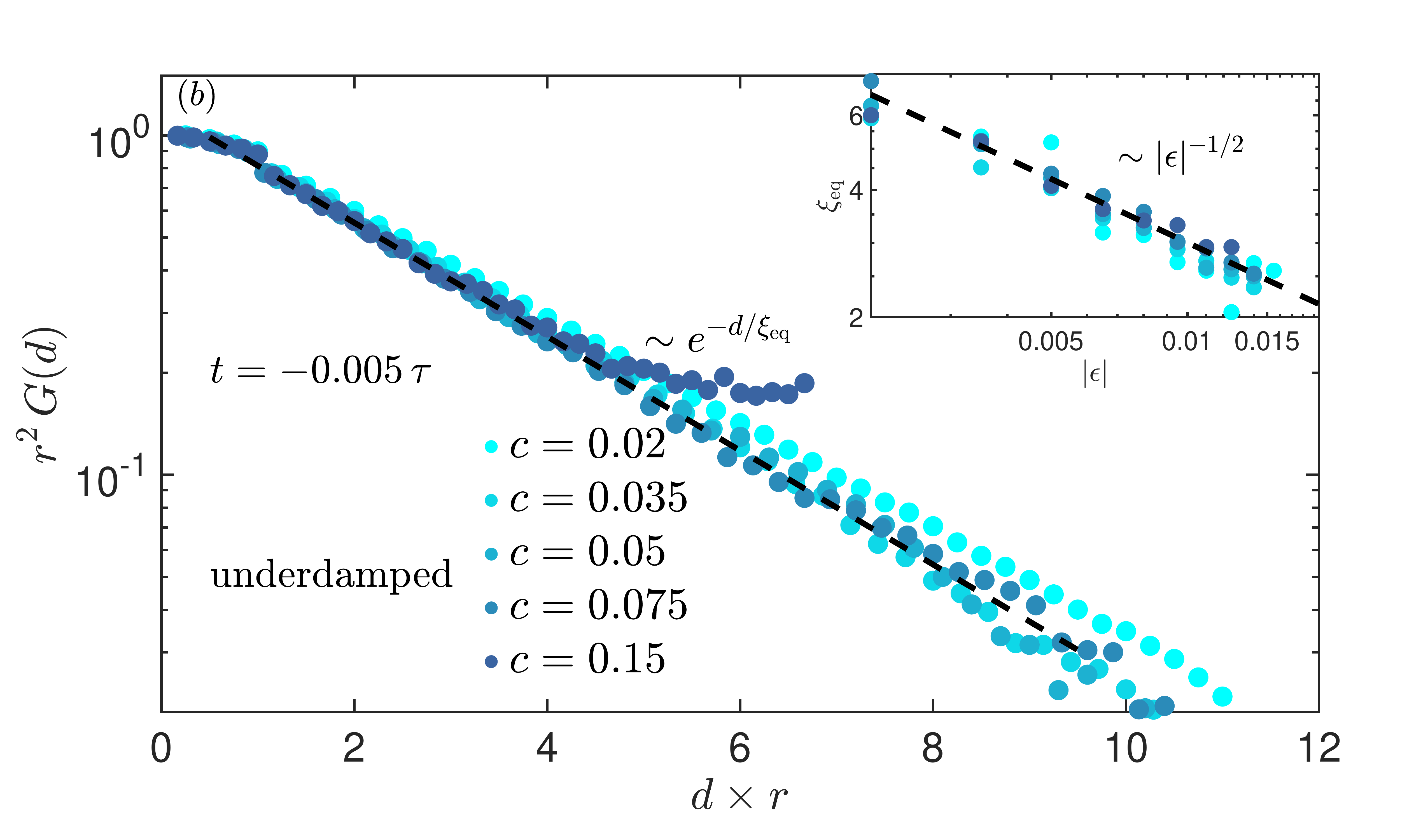}
\caption{Equilibrium correlation length for the overdamped and underdamped regimes, captured by the spatial and ensemble averages of the correlators of the $\sin\theta_i$ values for $N=401$. Panels show the scaling collapse for $(a)$ the overdamped and $(b)$ underdamped regimes via rescaling the distance by the regularity $r$, as motivated by the universal KZ limit. The inset displays the divergence of the correlation length as a function of the proximity to the critical point, quantified by $\epsilon$. The corresponding critical exponent is independent of $r$, up to numerical precision.  The results were obtained by averaging over $10^4$ trajectories and with quench time set to $\tau_Q=2560$ to ensure adiabatic time-evolution.\label{fig:Corr_r}}
\end{figure}

\begin{figure*}
\includegraphics[width=.95\linewidth,trim={0 0 0 0},clip]{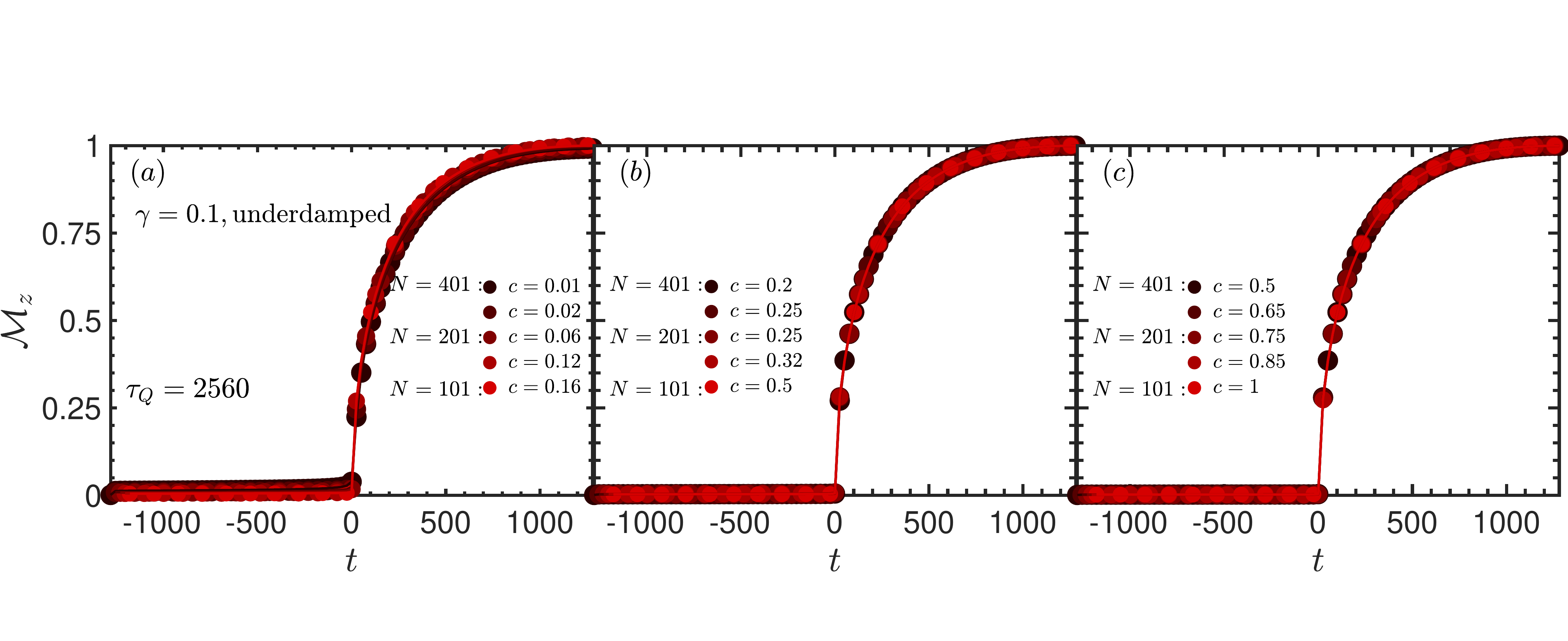}
\caption{\label{fig:order} Evolution of the average magnetization $\mathcal M_z$ as the order parameter for different values of $c$  in the $(a)$ small, $(b)$ intermediate, and $(c)$ fully connected limits for various system sizes. The first two pairs of $c$ values in the legends correspond to $N=401$ and $N=201$ sizes, respectively, while the last one is for $N=101$. The equilibrium dependence on the control parameter was extracted from time evolutions in the adiabatic limit, with  $\tau_Q=2560$ and averaged over $10^4$ trajectories. The critical properties observed in the chain topology are invariant with respect to $c$, maintaining the behavior around the critical point.}
\end{figure*}
Next, the equilibrium correlation length is investigated via the angle-angle correlation function defined as
\begin{equation}
G(d)=\overline{\langle\sin\theta_i\sin\theta_{i+d}\rangle_\mathrm{traj}}\approx\overline{\langle\theta_i\theta_{i+d}\rangle_\mathrm{traj}},
\end{equation}
where the last approximation is valid near the critical point. Here, the expressions $\langle\cdots\rangle_\mathrm{traj}$ and $\overline{\cdots}$ stand for the average over the stochastic trajectories and the spatial coordinates, respectively. As shown in App.~\ref{app:corr_length_r}, this allows for a compact analytical treatment in terms of a Hamiltonian quadratic in the angle variables $\theta_i$, yielding for the correlator
\begin{equation}
    G(d)\sim r^{-2}e^{-d/\xi_\mathrm{eq}},\quad \xi_\mathrm{eq}(r)\sim r^{3/2}\epsilon^{-1/2}.
\end{equation}
This result also implies the universal scaling of the correlation function as 
\begin{equation}
    r^2G(d)\sim e^{-d/\xi_\mathrm{eq}}.
\end{equation}
In contrast to the SMVC model, the SVL method provides access to the numerical investigation of equilibrium properties. This is achieved through an adiabatically slow time evolution, allowing the system to reach thermal equilibrium at temperature $T$ at all time instances via noise-induced fluctuations. To this end, the correlation function is computed by averaging over $3\times 10^4$ trajectories for $\tau_Q=2560$. This value was validated by numerically checking the convergence of the equilibrium quantities against increasing further the annealing time for $r=1$ and $N=401$, for which the strongest non-adiabatic effects are expected. Additionally, as it will be shown in Sec.~\ref{sec:KZM}, the final excess energy density reaches the onset of adiabaticity for smaller values of $\tau_Q$ for the investigated parameters, while its time-evolution matches precisely the analytical result in Eq.~\eqref{eq: E_min_t}.
For both damping regimes, the rescaled correlators exhibit an accurate scaling collapse as a function of $d/\xi_\mathrm{eq}$, while the correlation length grows approximately as $\xi_\mathrm{eq}\sim\lvert \epsilon\rvert^{-1/2}$ near the critical point, indicating $\nu=1/2$. The results of this numerical analysis are displayed in Fig.~\ref{fig:Corr_r}.

Finally, the effect of the finite-range interaction on the order parameter is put to the test. In analogy with the $r=1$ case, the classical version of the magnetization is introduced as the order parameter. This measure accurately quantifies the distance of the final rotor configuration from the ferromagnetic ground state,
\begin{equation}
    \mathcal{M}_z(t)=\frac{1}{N}\sum_{i=1}^N\left\langle|\sin\theta_i(t)|\right\rangle.
\end{equation} 
The magnetization is determined for the same value of $\tau_Q=2560$ and number of trajectories.
Remarkably, the sudden increase of $\mathcal M_z$ around $t_c=0$ follows the same shape regardless of $r$ and $N$.
The results are displayed in Fig.~\ref{fig:order} for the underdamped case in the regimes of small, intermediate, and large values of $c$.

In short, the SVL model captures the equilibrium properties of classical ferromagnetic $O(2)$ rotor network, indicating a continuous phase transition in which the critical properties are robust against variations of the interaction range, as long as the fundamental symmetries are not altered compared to the chain case. 

%
%
%
%
\section{\label{sec:KZM} Universal Kibble-Zurek scaling regime and its breakdown}

This section investigates how the excess energy density and the number of near-ferromagnetic domains are related to the quench time $\tau_Q$ and the connectance $c$.

\subsection{\label{subsec:KZM}The Kibble-Zurek mechanism}

The KZ mechanism predicts a universal power-law scaling of the density of excitations or topological defects generated while crossing a phase transition point \cite{Kibble_1976, Kibble_1980, Zurek_1985}. The critical slowing down prevents the system from adjusting to the driving-induced changes, leading to the formation of topological defects in the final state. 
To estimate the defect density, the KZ mechanism utilizes equilibrium scaling theory, characterizing the power-law divergence of the correlation length $\xi$ and the relaxation time $\tau$, 
\begin{equation}\label{eq:scalings}
    \xi=\xi_0|\epsilon|^{-\nu}\quad\text{and}\quad \tau=\tau_0|\epsilon|^{-z\nu},
\end{equation}
in the proximity of the transition. Here, $\nu$ and $z$ denote respectively the correlation-length critical exponent and the dynamic critical exponent, with $\tau_0$ and $\xi_0$ fixing the model-dependent time and length scales. 
Matching the driving rate and the relaxation time provides the freeze-out time-scale $\hat t$. Inside the corresponding freeze-out regime, $t\in[-\hat t,\hat t]$ dynamics ceases to be adiabatic. This sets the associated freeze-out length scale $\hat{\xi}$, determining the typical length scale of the domains
\begin{equation}\label{eq: KZ_rate}
\begin{split}
    &\epsilon(\hat t\,)/\dot\epsilon(\hat t\,)=\tau_0\lvert\epsilon(\hat t\,)\rvert^{-z\nu}\Rightarrow\hat t=\tau^{\frac{1}{z\nu+1}}_0\tau^{\frac{z\nu}{z\nu+1}}_Q\,,\\
    &\hat\xi=\xi_0\lvert\epsilon(\hat t\,)\rvert^{-\nu}.
    \end{split}
\end{equation}
Correspondingly, defect-free domains will extend over regions of size $\hat\xi^{d-D}$ leading to the average number of defects $\langle n\rangle$ scaling as
\begin{equation}
     \langle n\rangle\propto\hat\xi^{-(d-D)}\propto\xi^{-(d-D)}_0\left(\frac{\tau_0}{\tau_Q}\right)^{\frac{(d-D)\nu}{z\nu+1}},
\end{equation}
valid for defects extending along $D$ dimensions, when the spatial dimension of the system is $d>D$~\cite{del_Campo_2014}.

\subsection{Connectance dependence of the Kibble-Zurek scaling}

As in long-range quantum Ising models~\cite{Jaschke17}, defects arising in circulant rotor networks preserve their point-like or zero-dimensional character and are thus $\mathbb{Z}_2$ kinks with $D=0$. Within the domains, the rotors are approximately aligned ferromagnetically
with $\theta_i\approx\pm\pi$ and suddenly change their orientation at the location of a kink. 
This allows for a one-dimensional representation of the defects, the density of which can be estimated via the projected rotor values as
\begin{equation}\label{eq:kink_1}
    n_1=\frac{\mathcal{N}_1}{N} = \frac{1}{2N} \sum_{i=1}^{N-1} [1-\text{sgn}(\sin\theta_i)\text{sgn}(\sin\theta_{i+1})].
\end{equation}
According to the numerical results, this ``one-dimensional'' defect density follows precisely the same power law as in the chain topology, sufficiently below the fully connected limit $c\lesssim0.15$,
\begin{align}
    &n_1\sim\tau^{-1/4}_Q,\quad\text{ overdamped},\\
    &n_1\sim\tau^{-1/3}_Q,\quad\text{ underdamped}.
\end{align}
Even though the circulant graph topology does not modify the power law, it plays a crucial role in the universal properties of the defect density beyond rescaling the driving rate as $\sim r/\tau_Q$.

Motivated by these results we apply the framework of the KZ mechanism as an ansatz to provide an approximate analytical understanding of further dynamical quantities. Substituting the control parameter $\epsilon(t)=2r\,t/\tau_Q$ in the KZ rate equations~\eqref{eq: KZ_rate} yields the following freeze-out scales and the density of defects, 
\begin{equation}\label{eq:n_1KZM}
\begin{split}
    &\hat t=\tau^{\frac{1}{z\nu+1}}_0\left(\frac{\tau_Q}{r}\right)^{\frac{z\nu}{z\nu+1}}\,,\\
    &\hat\xi=\xi_0|\epsilon\big(\hat t\big)|^{-\nu}=\xi_0\left(\frac{\tau_Q}{\tau_0r}\right)^{\frac{\nu}{z\nu+1}}\,,\\
    &n_1\propto\hat\xi^{-(d-D)}\propto\xi^{-(d-D)}_0\left(\frac{\tau_0r}{\tau_Q}\right)^{\frac{(d-D)\nu}{z\nu+1}}.
\end{split}
\end{equation}
The identification of defects as
kinks and the one-dimensional topology of the 
resulting ferromagnetic domains imply that $d= 1$ and
$D= 0$.
The critical exponents $\nu,\,z$ and the microscopic constants $\tau_0,\xi_0$ may generally depend on the regularity in the KZ power-law regime.
 As shown in Eq.~\eqref{eq:overdamped} and Eq.~\eqref{eq:underdamped} and in App.~\ref{app:corr_length_r} both $z\nu$  and $\nu$ are independent of $r$. In particular, $z=1,\,2$ a for the overdamped and underdamped regimes, respectively, while $\nu=1/2,\,1/2$. As for the model-dependent constants, the regularity only enters via  $\xi_0\sim r^{3/2}$, while the relaxation time is independent of the graph topology, $\tau_0(r)\sim r^0$, see also Eqs.~\eqref{eq:overdamped} and ~\eqref{eq:underdamped}. 

Thus, the KZ exponents preserve their one-dimensional values, $\alpha_\mathrm{KZ}=1/4$ and $\alpha_\mathrm{KZ}=1/3$ for the overdamped and underdamped regimes, respectively. The KZ scalings, however, will not follow a universal function of $c$ due to the explicit $r$ dependence of the correlation length and the control parameter near the critical point. This leads to the results
\begin{align}\label{eq: n_1_averages}
    &\hat\xi\propto (r^5\tau_Q)^{1/4}\Rightarrow n_1\propto (r^5\tau_Q)^{-1/4},\,\text{ overdamped},\\
&\hat\xi \propto N(r^{7/2}\tau_Q)^{1/3}\Rightarrow n_1\propto (r^{7/2}\tau_Q)^{-1/3},\,\text{ underdamped}.
\end{align} 
As presented in Fig.~\figpanel{fig:averages_overdamped}{a} for the overdamped case and in Fig.~\figpanel{fig:averages_underdamped}{b} in App.~\ref{app: underdamped_averages} for the underdamped regime, the curves of $n_1$ follow the analytical approximations within numerical precision and exhibiting a precise scaling collapse for system sizes, $N=100,200,400$.

In addition, the scaling of the freeze-out time scale also acquires a regularity dependence,
\begin{equation}\label{eq: t_hat}
     \hat t\propto\begin{cases}
     &\left(\frac{\tau_Q}{r}\right)^{1/2},\quad\text{overdamped},\\
     &\left(\frac{\tau_Q}{r}\right)^{1/3},\quad\text{underdamped}.
     \end{cases}
\end{equation}

\subsection{\label{sec:kinknumber} Universal energy density beyond nearest-neighbors}

To demonstrate the applicability of the KZ mechanism, we investigate the relationship between $n_1$ and excess energy production. The circulant graph topology disrupts the simple one-to-one correspondence between the number of near-ferromagnetic domains and the excess energy. All rotor configurations within the interaction range contribute to the latter, which is defined as follows
\begin{equation}\label{eq: Delta_E}
\begin{split}
    &\rho_E(t) = \frac{H(\bm\theta,t)-E_\mathrm{min}(\bm\theta,t)}{Nr},\\
    &\rho_E\equiv\rho_E(\tau_Q)= 1-\frac{1}{4}\sum_{i,j=1}^N A_{ij}\sin\theta_i\sin\theta_j.
    \end{split}
\end{equation}
Here, the first term denotes the time evolution of the excess energy density, while the second one is for its final value, which also takes a simple form in terms of the quadratic part of the Hamiltonian.

\begin{figure*}
    \includegraphics[width=.95\linewidth]{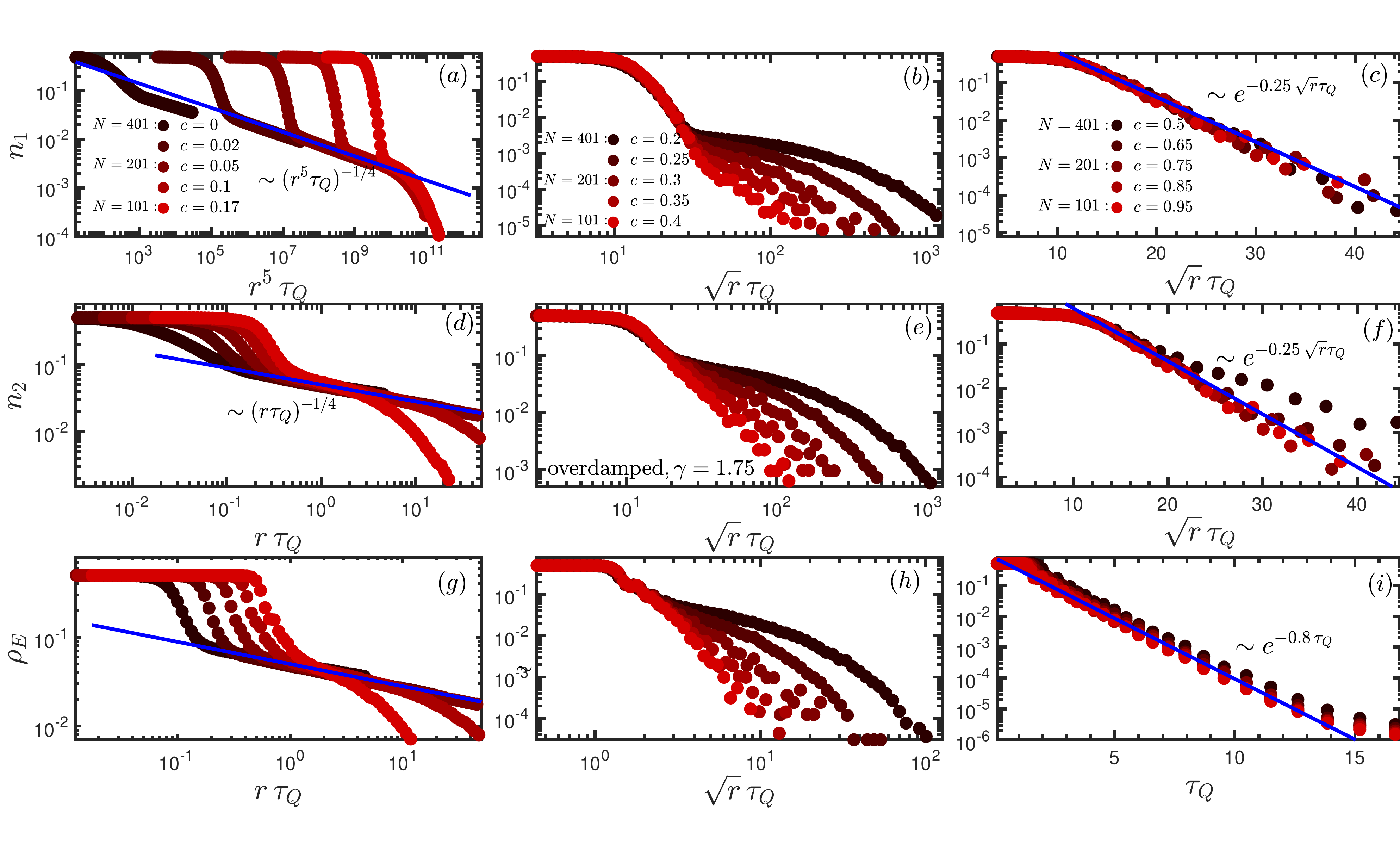}
    \caption{Averages of the defect and excess-energy densities as functions of $\tau_Q$ in the overdamped regime with the connectance varying from the chain topology towards the fully connected limit. The first two pair of $c$ values in the legends correspond to $N=401$ and $N=201$ sizes, respectively, while the last to $N=101$.
    Panels $(a)-(c)$: One-dimensional defect density. For values $c\lesssim0.15$, the same power law of the chain topology is observed with a universal regularity rescaling. For intermediate and densely connected graphs, a crossover is displayed towards the adiabatic regime, exhibiting a universal exponential shape. 
    Panels $(d)-(f)$: The graph defect density  $n_2$ behaves similarly,  exhibiting power laws of $r$ different from the KZ regime but identical exponential behavior in the adiabatic regimes.
    Panels $(g)-(i)$: The excess energy density closely matches the behavior of graph defect density with high precision, including additional features in the absence of defects that stem from the dynamics of small angle deviations around the ferromagnetic direction. In all regimes, the curves were averaged over $10^3$ trajectories.}
    \label{fig:averages_overdamped}    
\end{figure*}
Here, the first equation accounts for the excess energy during the evolution, while the second is for its density at the end of the schedule.
The normalization in the second row makes the density scale invariant with respect to both $N$ and $r$.
Here, the minimum of the instantaneous ground state energy $E_\mathrm{min}(\bm\theta,t)=\min_{\{\theta_i\}_{i\in V}} H(\bm\theta,t)$ is defined as the global minimum of the Hamiltonian at time $t$.
By translational invariance, all angles are identical, and the search for the instantaneous energy minimum simplifies to the minimization of a single-variable function. For convenience, the schedule is considered in the interval $[0,\tau_Q]$ with the critical time given by $t_c=\tau_Q/2$,
\begin{equation}
\begin{split}
     \partial_\theta H(\theta,t)&=rN\partial_\theta \left[-\frac{t}{\tau_Q}\sin^2\theta-2\left(1-\frac{t}{\tau_Q}\right)\cos\theta\right]=0\\
     &\Rightarrow \frac{t}{\tau_Q}\sin\theta\cos\theta=\left(1-\frac{t}{\tau_Q}\right)\sin\theta.
     \end{split}
\end{equation}
For $t\leq\tau_Q/2$ this equation has the trivial solution $\theta=0$ and with minimum energy $E_\mathrm{min}(\bm\theta,t)=-2rN\left(1-t/\tau_Q\right)$, while for $t>\tau_Q/2$ the solution is given by $\cos\theta=\frac{\tau_Q-t}{t}$. Thus, the minimum energy yields
\begin{equation}\label{eq: E_min_t}
\begin{split}
    E_\mathrm{min}(\bm\theta,t)=-rN\biggl[\frac{t}{\tau_Q}\biggl[1&-\left(\frac{\tau_Q-t}{t}\right)^2~\biggr]
    \\&+2\frac{\tau_Q}{t}\left(1-\frac{t}{\tau_Q}\right)^2~\biggr].
\end{split}
\end{equation}
This expression provides a convenient way for the numerical computation of Eq.~\eqref{eq: Delta_E}.
The excess energy is set by the actual rotor configuration with contributions attributed to the kinks as well as angle deviations within the domains. 
However, these angle fluctuations provide negligibly small contributions compared to the energy scale of the kinks, and $\rho_E$ to leading order can be expressed in terms of $n_1$. For values of $\tau_Q$ where the freeze-out length sufficiently exceeds the interaction range ($\hat\xi>r$), every defect carries an energy contribution of $\sim r^2$. This originates from the antiparallel rotors on opposite sides of the defect.
To this end, $n_1$ is also generalized by counting all rotor-rotor fluctuations within the interaction range, introducing the graph defect density
\begin{equation}\label{eq:kink2}
    n_2=\frac{\mathcal{N}_2}{Nr} = \frac{1}{4Nr} \sum_{i,j = 1}^N A_{ij}[1-\text{sgn}(\sin\theta_i)\text{sgn}(\sin\theta_j)].
\end{equation}
Within the domain of the KZ scaling, the graph defect density agrees well with the numerically exact excess energy density. Due to the normalization with the connectivity, $2rN$, both $\rho_E$ and $n_2$ are also proportional to $n_1$ by a factor of $r$.
Remarkably, $\rho_E$ and $n_2$ follow the same universal function of $\tau_Q/\hat t(r)$ independently of both $r,\,N$ and the respective damping regimes up to the leading order,
\begin{equation}
    \begin{split}
        &n_2\propto r n_1\propto\left(\frac{\tau_Q}{\hat t(r)}\right)^{-1/2},\\
        &\rho_E=\frac{\Delta E}{Nr}\propto n_2\propto
        \begin{cases}
        &(r\,\tau_Q)^{-1/4},\quad\text{overdamped},\\
        &(\sqrt r\,\tau_Q)^{-1/3},\,\text{underdamped},
        \end{cases}
    \end{split}
\end{equation}
as displayed in Fig.~\figpanel{fig:averages_overdamped}{d,g} for the overdamped case for different values of $N$ and $c$ (see also Fig.~\figpanel{fig:averages_underdamped}{d,g} in App.~\ref{app: underdamped_averages} for the underdamped case).


\section{Breakdown of Kibble-Zurek mechanism and the fully connected limit}\label{SecKZMbreaks}

In this section, the breakdown of the universal KZ scaling regime is investigated in terms of the regularity $r$, connectance $c$, and quench time $\tau_Q$. Subsequently, the universal signatures of annealing dynamics near the fully connected limit are explored. Finally, universal properties of $\rho_E(t)$ are analyzed at intermediate times, $t\in[-\tau_Q/2,\tau_Q/2]$ in both the KZ and the adiabatic regimes.

\subsection{Fast quenches}
As in the transverse field Ising model (TFIM) and the $\phi^4$ classical stochastic model ~\cite{zeng_universal_2023,Han_Chaun_2023_Fast}, the universal power-law scaling of both $n_1$ and $\rho_E$ breaks down for rapid annealing processes.
For small $\tau_Q$, the condition $\hat\xi<r$ coincides with the fast quench breakdown of the KZ scaling.
In particular, the $\tau_Q/r$ rescaling in the control parameter
allows the system to stay inside the freeze-out regime for the whole process. The fast quench breakdown can be captured by matching the freeze-out time with the equilibrium relaxation time, $\hat t=\tau(\tau_Q)$~\cite{zeng_universal_2023} at the end of the process. It sets the condition for $\tau_Q$ and the lower threshold of the freeze-out length scale
\begin{equation}\label{eq:fast_condition}
\begin{split}
    &\left(\frac{\tau_Q}{r}\right)^{\frac{z\nu}{z\nu+1}}=\tau_0r^{-z\nu}\Rightarrow \tau^{\mathrm{fast}}_Q\propto r^{-z\nu},\\
    &\Rightarrow\hat\xi\propto r^{3/2}\left(\frac{\tau_Q}{r}\right)^{\frac{\nu}{z\nu+1}}\propto r^{3/2-\nu}= r,
    \end{split}
\end{equation}
using $\nu(r)=1/2$ in the last step. Thus, the fast quench analysis via matching the relaxation and freeze-out time scales provides the same result as using the condition of $\hat\xi=r$,
\begin{equation}
    \tau^\mathrm{fast}_Q\sim r^{-z\nu}\sim
    \begin{cases}
    &r^{-1},\,\,\,\,\,\quad\text{overdamped},\\
    &r^{-1/2},\quad\text{underdamped}.
    \end{cases}
\end{equation}

Furthermore, this approach is validated by the fact that by increasing either $r$ or $\tau_Q$, the freeze-out time scale remains inside the domain of validity of the linear approximation, $\hat t/\tau_Q\ll1$. Thus, increasing the regularity decreases the critical annealing time below which the defect and excess energy densities exhibit a fast quench plateau, as demonstrated in Fig.~\figpanel{fig:averages_overdamped}{a,d,g} for $n_1,\,n_2$ and $\rho_E$, respectively. As $r$ is increased, these measures terminate earlier at constant plateaus attributed to the fast quench regime (see also App.~\ref{app: underdamped_averages} for the underdamped case). However, for larger values of $c$, these plateaus disappear in the limit $N\rightarrow\infty$.

\subsection{Fully connected limit and adiabatic regime}\label{sec: LMG}

As discussed in Sec.~\ref{sec:kinknumber}, considering the similarities between the one-dimensional rotor chain and the TFIM model in both damping regimes, one could expect the complete rotor network with graph $K_N$ to parallel the behavior of the Lipkin-Meshkov-Glick (LMG)  model \cite{Caneva08}. However, the results clearly show that this is not the case.

Increasing $\tau_Q$ naturally leads to adiabatic dynamics; however, for rotor networks, the approach to the adiabatic limit also strongly depends on the regularity of the graph.
The KZ regime only persists until the onset of adiabaticity identified by the condition $n_1=\frac{1}{N}$. 
This sets the upper threshold of of $\tau_Q$ in terms of $r$ and $N$
\begin{align}\label{eq:ad_condition}
    &r^{-3/2}\left(\frac{\tau_Q}{r}\right)^{-\frac{\nu}{z\nu+1}}\propto\frac{1}{N}\\
    &\Rightarrow\tau^\mathrm{ad}_Q\propto N^4r^{-5}\quad\text{and}\quad \propto N^3r^{-7/2},
\end{align}
for the overdamped and the underdamped regimes, respectively. Note that this condition is equivalent to matching the freeze-out length scale with the system size, $\hat \xi=N$.
 
Thus, the universal KZ power-law regime completely disappears around the threshold value of the connectance for which the fast quench time scale~\eqref{eq:fast_condition} and the adiabatic one in ~\eqref{eq:ad_condition} become of the same order in the limit $N,r\rightarrow\infty$. This leads to the universal breakdown scale of
\begin{equation}\label{eq:KZM_disappear_scale}
    \tau^\mathrm{fast}_Q=\tau^\mathrm{ad}_Q\Rightarrow r\sim N\rightarrow c\sim\mathrm O(1),
\end{equation}
for both the underdamped and overdamped regimes.
The breakdown scales also suggest that the larger $r$ is, the earlier the KZ regime sets in. However, as $N$ increases, the adiabatic regime takes over the dynamics even faster, eventually washing out the power-law KZ scaling.
Remarkably, the corresponding threshold connectance remains finite in the thermodynamic limit, with $ c\approx0.15$ for both damping regimes. 
When the connectance $c\in[0.15,0.5]$, the KZ regime disappears, and the defect densities exhibit finite fast quench plateaus followed by an intermediate non-universal sharply decaying regime. The corresponding shrinking and vanishing of the universal KZ regime are displayed in
Fig.~\figpanel{fig:averages_overdamped}{b,c} for $n_1$ and Fig.~\figpanel{fig:averages_overdamped}{e,f,h,i} for $n_2$ and $\rho_E$ in the overdmaped regime (see also Fig.~\ref{fig:averages_underdamped} in App.~\ref{app: underdamped_averages} for the underdamped regimes). 

Beyond this scale, $c\gtrsim0.5$, the defect and energy densities turn into a sharp, exponential decay.
This regime exhibits new distinct universal behavior taking over the dynamics, with the fast quench plateaus disappearing with increasing $N$. In both cases, the universal regime also becomes independent of $N$.

The exponential decay of $n_1$ and $n_2$ is shown in Fig.~\figpanel{fig:averages_overdamped}{c,f,i} (see also Fig.~\figpanel{fig:averages_underdamped}{c,f,i} in App.~\ref{app: underdamped_averages}), for quench times, where the last effect pair can still be detected numerically for the used number of stochastic trajectories. These curves become universal as a function of the rescaled annealing time $r^{\frac{z\nu}{z\nu+1}}\tau_Q$,
  \begin{equation}\label{eq: n_1_n_2_exp_decay}
  \begin{split}
      &\log n_1\sim\log n_2\sim-0.25\,r^{1/2}\tau_Q,\quad\text{overdamped},\\
      &\log n_1\sim\log n_2\sim-0.7\,r^{1/3}\tau_Q,\quad\text{underdamped}.
      \end{split}
  \end{equation}
Although the decaying part exhibits non-universal behavior in the intermediate regime $c\in[0.15,0.5]$, the best scaling collapse for moderate values of $\tau_Q\lesssim10$ is obtained by the same rescaling as in Eq.~\eqref{eq: n_1_n_2_exp_decay}. This is shown in Fig. \figpanel{fig:averages_overdamped}{b,d} in the overdamped regime, for $n_1$ and $n_2$, respectively (see also App.~\ref{app: underdamped_averages} for the underdamped case).

  By contrast, the excess energy can take finite values also for much larger values of $\tau_Q$ due to small angle deviations compared to the $\theta=\pi/2$ ferromagnetic direction. Within this defect-free regime, $\rho_E$ is governed by these small angle fluctuations. 
  Its behavior thus differs from that of $n_1$ and $n_2$. 
  As shown in Fig.~\figpanel{fig:averages_overdamped}{i} for the overdamped regime (see also App.~\ref{app: underdamped_averages} for the underdamped case), the exponential decay of the excess energy becomes independent of $r$ up to exponential accuracy and follows the curve given by the best numerical fit,
\begin{equation}
    \begin{aligned}
      &\log\rho_E\sim\,0.8\tau_Q,\quad\text{overdamped},\\
      &\log\rho_E\sim\,0.06\tau_Q,\,\,\,\text{underdamped}.
    \end{aligned}
\end{equation}
As for the intermediate regime $c\in[0.15,0.5]$, the best scaling collapse for moderate values of $\tau_Q$ is achieved by a slightly different rescaling of $\tau_Q\rightarrow\sqrt r\,\tau_Q$, as observed in Fig.~\figpanel{fig:averages_overdamped}{h} for the overdamped regime. Quite remarkably, in the underdamped regime, a perfect scaling collapse is observed independently of $r$, as shown in App.~\ref{app: underdamped_averages}.

\subsection{Universal breakdown: general case}\label{sec:General_breakdown}
Next, we show that the finite-range extensions of arbitrary one-dimensional systems critical exponents $z$ and $\nu$  exhibit the same universal breakdown of the KZ power-law scaling regime as in the case of $O(2)$ rotors.
Consider a general Hamiltonian, 
\begin{align}
    &H(t)=-\frac{J(t)}{2}\,H_P-h(t)H_0,\\
    &H_P=\sum_{i,j}A_{ij}\mathcal O_i\mathcal O_j,\\
    &H_0=\sum_i\varepsilon_i.
\end{align}
where $A_{ij}$ denotes the adjacency matrix of a circulant graph with interaction range $r$, as defined in Eq.~\eqref{eq: A_ij_def}. Here, $\mathcal O_i$ are either quantum mechanical operators or classical variables accounting for the generalized interaction Hamiltonian. The operators or classical variables $\varepsilon_i$ stand for the generalized transverse fields. We assume that these operators are such that in the one-dimensional chain topology, i.e., for $r=1$, the system governed by $H(t)$ exhibits either a quantum or a classical phase transition at $h(t_c)=J(t_c)$ with critical exponents $z$ and $\nu$. Prominent examples of such Hamiltonians are both classical and quantum Ising models with spins of arbitrary length, interacting bosons, or the fermionic/bosonic Fermi-Hubbard model.~\cite{Sachdev_2011}

As in the case of $O(2)$ rotors, for the universal KZ scaling to hold and for its breakdown to be universal, we consider that the equilibrium correlation length and relaxation time preserve a power law scaling over a range of connectance values.  This preserves the notion of locality, the dimensionality of the system, and the defects, i.e., $d=1,\,D=0$.
The relaxation time scale coefficient $\tau_0$ is also considered to be independent of $r$; see App. \ref{AppC}. In the classical case, one can repeat the derivation in Eq.~\eqref{eq:overdamped} and Eq.~\eqref{eq:underdamped}, where the interaction range only enters in the additional $r$ scale of the generalized transverse field strength, associated with the terms $\epsilon_i$.
More generally, when interactions still act locally, $r$ only varies the energy scale of $H_P$ linearly in the proximity of the critical point. Thus, only the critical point will be shifted, leaving the power-law divergence of relaxation time intact, $\tau_0\sim r^0$. The same rescaling of $h(t)$ applies, $h(t)=r\left(1-\frac{t}{\tau_Q}\right),\,J(t)=\frac{t}{\tau_Q}$ as in the case of rotors. As a result, the effective control parameter acquires the same linear scaling with respect to $r$ near the critical point, $\epsilon(t)\sim\frac{rt}{\tau_Q}$ and $\tau\sim\lvert\epsilon\rvert^{-z\nu}$.
As far as the correlation length is concerned, a general argument can be applied similar to that of the rotors. In the same domain of validity of locality the correlation length scales linearly
with $r$ near the phase transition, with the same critical exponent $\nu$.
Thus, with the rescaled control parameter, one arrives at the general critical behavior of $\xi\sim r^{1+\nu}\lvert \epsilon\rvert^{-\nu}$.

With these considerations, applying the KZ rate equation with the rescaled control parameter, one arrives at
\begin{equation}\label{eq:n_1KZM}
\begin{split}
    &\hat t=\tau^{\frac{1}{z\nu+1}}_0\left(\frac{\tau_Q}{r}\right)^{\frac{z\nu}{z\nu+1}}\,,\\
    &\hat\xi=\xi_0|\epsilon\big(\hat t\big)|^{-\nu}=\xi_0\left(\frac{\tau_Q}{\tau_0r}\right)^{\frac{\nu}{z\nu+1}}\,,\\
    &n_1\propto\hat\xi^{-1}\propto\xi^{-1}_0\left(\frac{\tau_0r}{\tau_Q}\right)^{\frac{\nu}{z\nu+1}}\\
    &\quad\sim \tau^{-\frac{\nu}{z\nu+1}}_Q\,r^{-\frac{z\nu(\nu+1)+1}{z\nu+1}}.
\end{split}
\end{equation}
The fast quench breakdown is identified by the same relation as in Eq.~\eqref{eq:fast_condition}. The condition of $\hat t=\tau_Q$ leads to the same relation, $\tau^\mathrm{fast}_Q\sim r^{-z\nu}$. However, it is instructive to validate the dependence of $\xi_0$ and $\tau_0$ on $r$ by showing that the fast quench threshold can also be obtained by matching the freeze-out scale with the interaction range, $\hat\xi=r$, as in Eq.~\eqref{eq:fast_condition}. In particular,
\begin{equation}
\begin{split}
    \hat \xi&\sim r^{\nu+1}\left(\frac{\tau_Q}{r}\right)^{\frac{\nu}{z\nu+1}}\propto r\\
    &\Rightarrow \tau^\mathrm{fast}_Q\sim r^{\frac{z\nu+1}{\nu}+1-\frac{\nu+1}{\nu}(z\nu+1)}\sim r^{-z\nu},
    \end{split}
\end{equation}
matching the expected scale for arbitrary $z$ and $\nu$.

The onset of adiabaticity is again identified by $n_1\sim N^{-1}$, leading to the adiabatic time-scale
\begin{equation}
\begin{split}
    n_1&\sim \tau^{-\frac{\nu}{z\nu+1}}_Q\,r^{-\frac{z\nu(\nu+1)+1}{z\nu+1}}\sim N^{-1}\\
    &\Rightarrow\tau^\mathrm{ad}_Q\sim N^{\frac{z\nu+1}{\nu}}r^{-\frac{z\nu(\nu+1)+1}{\nu}}.
    \end{split}
\end{equation}
Finally, matching the fast and adiabatic time-scales one obtains for the breakdown of the KZ power-law regime,
\begin{equation}
\begin{split}
    \tau^\mathrm{fast}_Q&=\tau^\mathrm{ad}_Q\Rightarrow r^{-z\nu}\sim N^{\frac{z\nu+1}{\nu}}r^{-\frac{z\nu(\nu+1)+1}{\nu}}\\
    &\Rightarrow N\sim r,
    \end{split}
\end{equation}
implying that the KZ regime survives for a finite ratio of the interaction range and the system size, i.e., for finite connectance values, extending the universality of the KZ breakdown to arbitrary one-dimensional critical systems under the given assumptions.

\subsection{\label{sec:dynamics}Universal dynamics and scaling collapse}
So far, the defect and excess energy densities have been investigated at the end of the time evolution as a function of $r$ and $\tau_Q$. To provide a more complete picture of the emerging universal properties, dynamical characteristics at intermediate times during the quench are also put to the test.
\begin{figure}
\includegraphics[width=\linewidth]{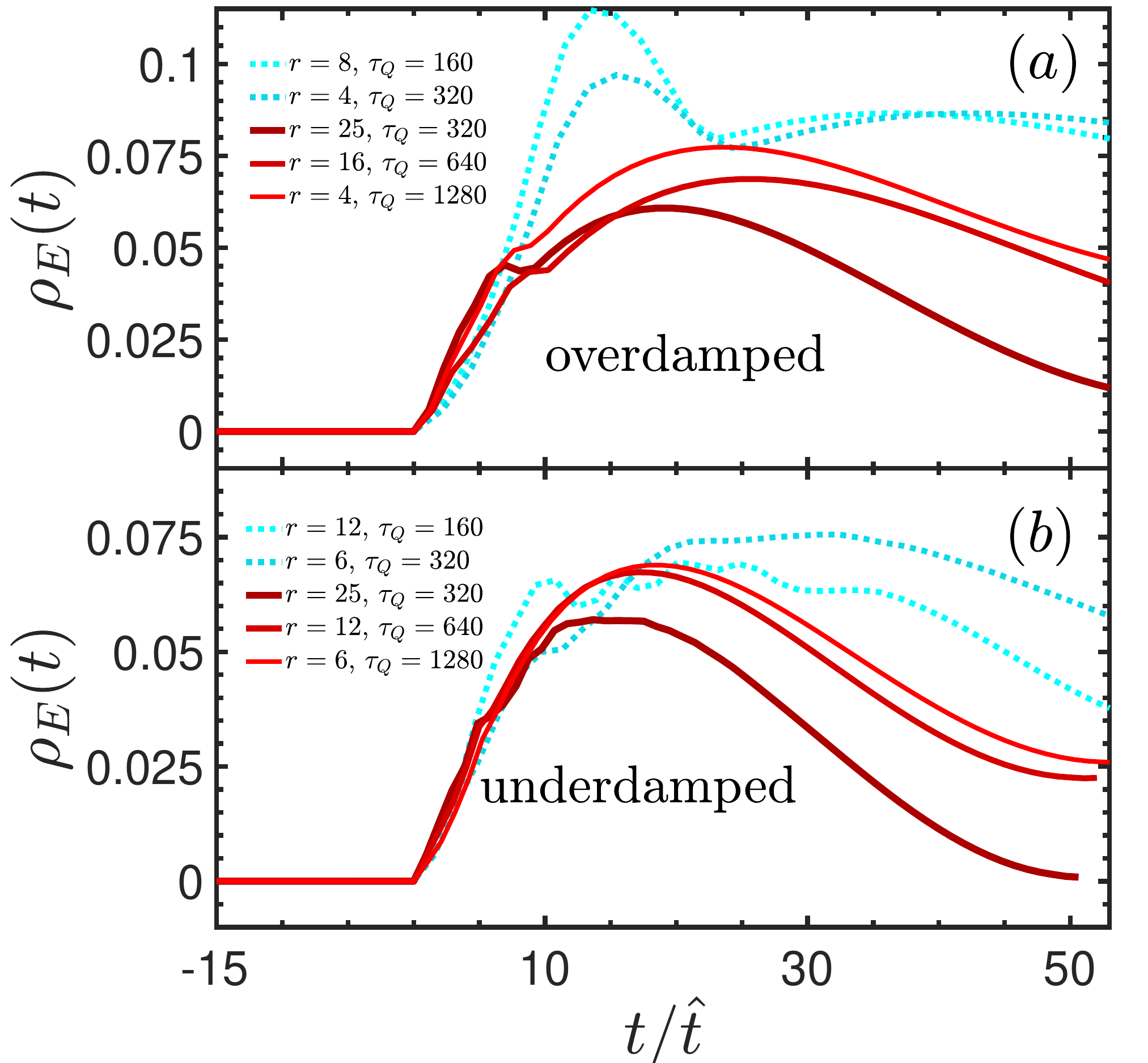}
\caption{Time evolution of the excess energy density near the critical point. The universal character of the dynamics is demonstrated for different values of $r,\,\tau_Q$ and $N$ by displaying $\rho_E(t)$ as a function of $t/\hat t$ in the KZ regime both for $(a)$  the overdamped and $(b)$ the underdamped regime. Blue symbols were obtained with $N=201$ while the red curves with $N=401$. In both cases, $10^4$ trajectories were employed.}\label{fig:t_t_hat}
\end{figure}
The time-evolution of $\rho_E(t)$ near the critical point is in good agreement with the predictions of the KZ mechanism.
The growth of $\rho_E(t)$ becomes independent of $r,\,\tau_Q$ and $N$ for connectances inside the KZ regime following a single universal curve as a function of $t/\hat t$ up to numerical precision. This is demonstrated in Fig.~\figpanel{fig:t_t_hat}{a,b} for the overdamped and underdamped regimes, respectively.

\begin{figure}
\centering
\includegraphics[width=1\columnwidth]{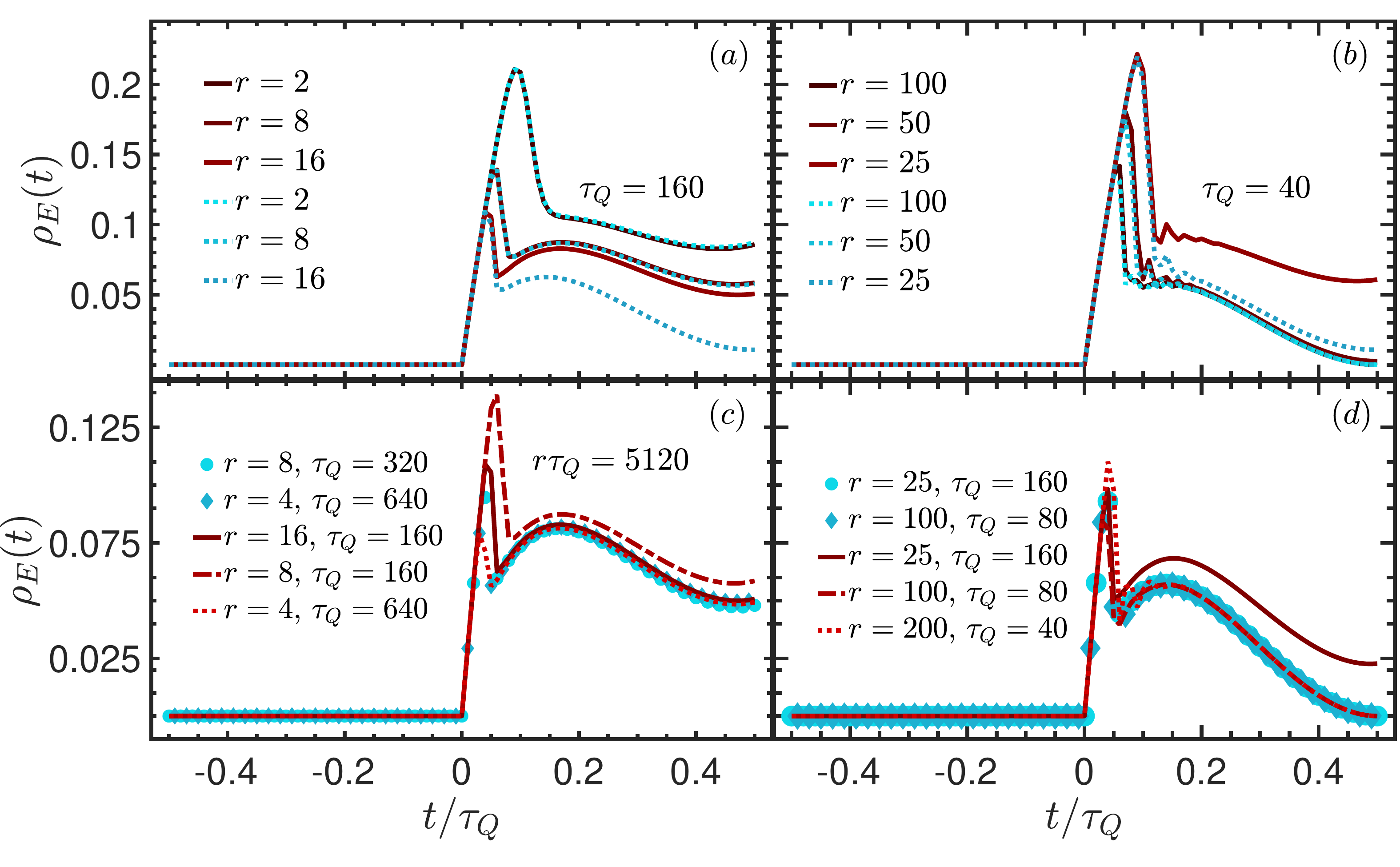}
\caption{\label{fig:dynamics} Time-evolution of the energy density $\rho_E(t)$  as a function of $t/\tau_Q$ for the overdamped regime for different values of $N,\,\tau_Q$ and $r$. Panels $(a),\,(b)$: Fixed quench times for pairs of $r$ and $N$ exhibiting size-independent behavior but becoming non-universal with respect to $r$ both in $(a)$  the KZ ($\tau_Q=40$) and  $(b)$ the adiabatic regimes ($\tau_Q=160$).
Panel $(c)$: Varying $\tau_Q$ with fixed $r\tau_Q=5120$. Good scaling collapse is found in the KZ regime for different values of $N$. Panel $(d)$: Adiabatic regime of $c\gtrsim0.5$ and $\tau_Q\gg\hat t$ featuring similar scaling collapse in the adiabatic regime. In all panels, averages involve $10^4$ trajectories, with blue symbols for $N=201$, and red lines for $N=401$.}
\end{figure}

Moreover, additional universal signatures emerge in both the KZ and adiabatic regimes with $r$ independent time-evolution along the whole process.
As shown in Fig.~\figpanel{fig:dynamics}{a,b}, the curves of $\rho_E(t)$ exhibit precise scaling collapse for different values of $N$. This universality only holds inside the KZ and adiabatic regimes, as also shown by the deviations of the curves with $N=200,\,r=12$ reaching the boundary of the KZ regime and $N=400,\,r=25$ near the onset of adiabaticity.
However, for fixed $\tau_Q$, the time-evolution exhibits non-universal behavior with respect to $r$ even
when plotted for the rescaled time, $t/\tau_Q$. As displayed in \figpanel{fig:dynamics}{a,b} by decreasing either $r$ or $\tau_Q$, $\rho_E(t)$ exhibits stronger non-adiabatic effects with larger final values and more pronounced peaks around the critical point. 
In this case, the final values follow either KZ power laws or decay exponentially in the adiabatic regime, depending on $r$ and $\tau_Q$.

We note that varying the annealing time while keeping fixed the ratio $\tau_Q/\hat t(r)$ in the KZ regime, or the product of $r^{\frac{z\nu}{1+z\nu}}\tau_Q$ in the adiabatic regime, unveils a precise scaling collapse of the date for different values of $N$ and $r$. This is demonstrated in Fig.~\figpanel{fig:dynamics}{c} in the KZ regime and in Fig.~\figpanel{fig:dynamics}{d} in the adiabatic case.  As with the dependence of $n_1$ and $\rho_E$ on $\tau_Q$,  the nonequilibrium universal dynamics is brought out upon separate rescaling by $N$ and $r$, rather than by only $c$. The proper rescaling can take different forms depending on the investigated quantities, the damping regimes, and the power-law or the exponential behavior.

\begin{figure*}[t!]
\includegraphics[width=\linewidth]{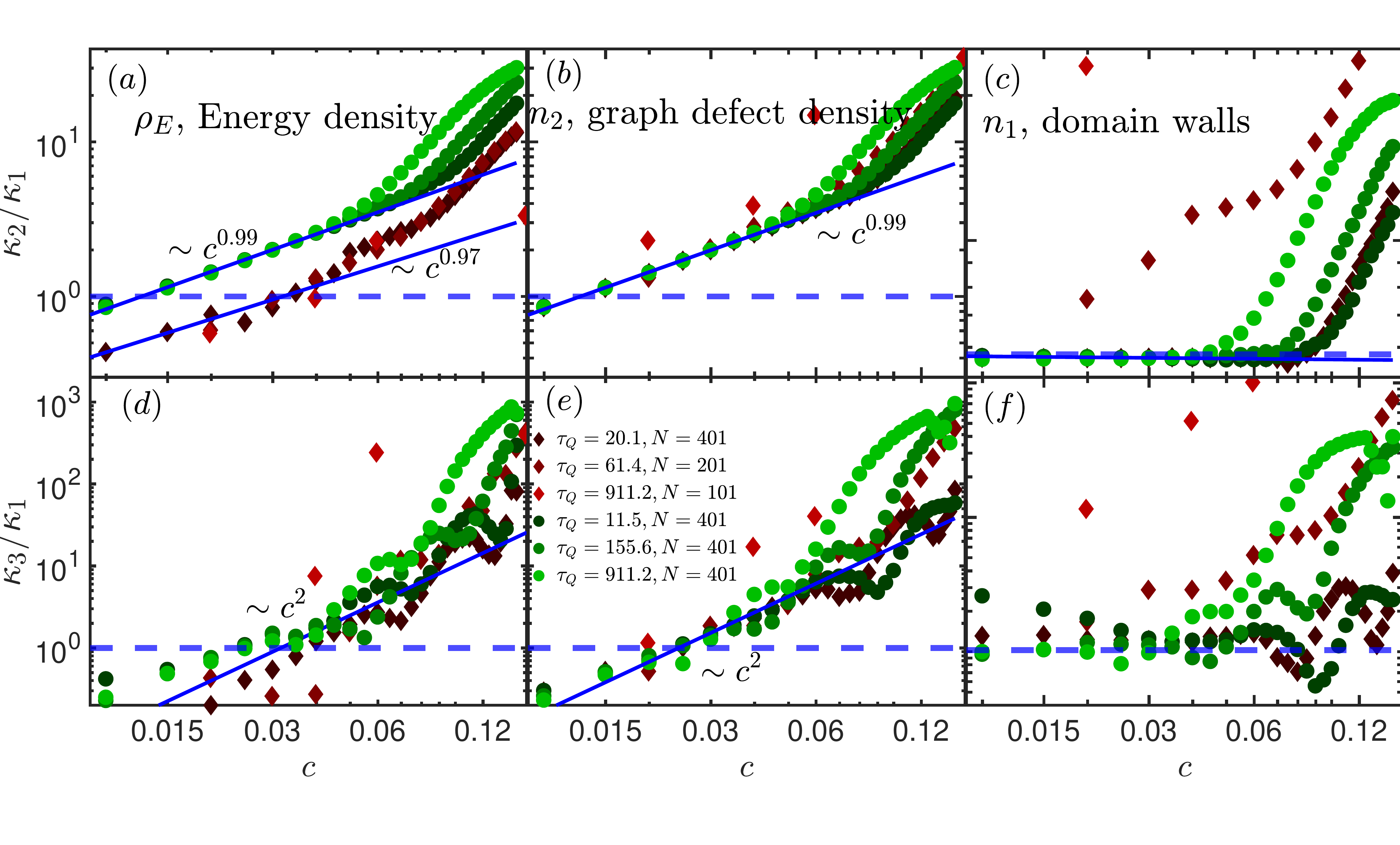}
\caption{\label{fig:cumulants_rho_E_n_2_n_1} Cumulant ratios  
as a function of $c$ for $n_1,\,n_2$ and $\rho_E$, for fixed values of $\tau_Q$ and for system sizes $N=401,\,201,\,101$.
Panels $(a)-(c)$: The ratio $\kappa_2/\kappa_1$ behaves similarly and exhibits the predicted universal linear growth for $\rho_E$ and $n_2$, while it remains constant for $n_1$. 
For relatively small values of $c$, a crossover is observed between the sub- and super Poissonian behavior within the KZ regime, while the same happens for larger values of $c$ in the case of the $n_1$ around the breakdown of the power-law regime. These cumulant ratios are estimated from  $5\times 10^4$ trajectories.
Panels $(d)-(f)$: The ratio $\kappa_3/\kappa_1$ exhibits a universal quadratic growth with $c$ inside the KZ regime for $\rho_E$ and $n_2$, while it follows the predicted constant value for $n_1$. Cumulants are extracted from  $10^4$ trajectories.}
\end{figure*}
%
%
\section{\label{sec:beyondKZM}Beyond the Kibble-Zurek mechanism: fluctuations in the defect and energy densities}
In this section, we investigate the second and third cumulants of the defect densities and excess energy.
Beyond the KZ scaling of the average defect densities, higher-order cumulants are of utmost importance regarding the full energy distribution.
In the quantum case, it was shown for the TFIM that the probability distribution to find $n$ kinks in the system at the end of the quench, $P(n)=\mathbb{E}[\delta(\mathcal{N}-n)]$, follows a Poisson binomial distribution \cite{delCampo18,Cui19}, as demonstrated in annealing devices \cite{Bando20,Bando21,King22}. In fully connected quantum systems, a negative binomial distribution of fractional
index was found instead \cite{Gherardini23}.
In the classical domain, a binomial distribution was predicted for point-like defects \cite{delCampo18,GomezRuiz19,Mayo21,Subires22,GomezRuiz22,Thudiyangal24}, with a Poisson binomial distribution occurring in the more general setting \cite{GomezRuiz19,Xia2024}. Fourier transforming $P(n)$ yields the characteristic function $\hat{P}(\phi)=\mathbb{E}[e^{i\phi n}]$, whose logarithm gives the cumulant generating function $\ln{\mathbb{E}[e^{i\phi n}]}=\sum_{p=1}^\infty\frac{\kappa_p}{p!}(i\phi)^p$, where $\kappa_p$ is the $p$-th order cumulant. It has been argued that all cumulants share the same power-law scaling exhibited by the mean, $\kappa_q \propto \kappa_1$ for every $q$. As a result,  the ratio $\kappa_q/\kappa_1$ is constant and independent of $\tau_Q$ \cite{delCampo18, GomezRuiz19}. For instance, for the TFIM $\kappa_2/\kappa_1=2-\sqrt{2}\approx0.578$ \cite{delCampo18,Cui20}, as verified in D-Wave devices \cite{Bando20}. As for classical systems, the same binomial relation of the first three cumulants has been shown for the chain topology~\cite{Subires22}. 
In what follows, we focus on the second and third cumulants, $\kappa_2$ and $\kappa_3$, beyond the average density ($\kappa_1$).

\begin{figure*}
    \includegraphics[width=.95\linewidth]{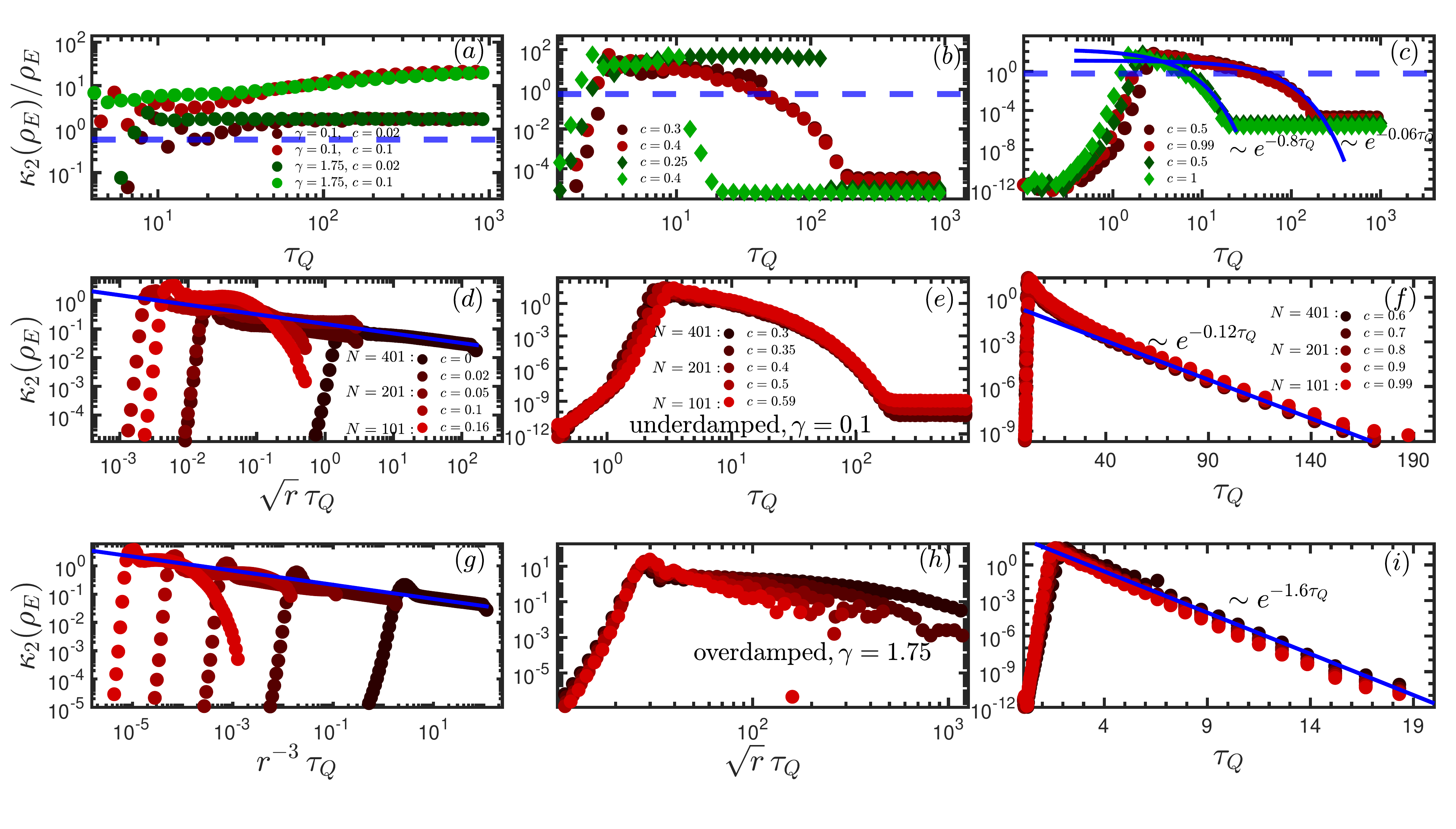}
    \caption{Second cumulants of the excess energy density as a function of $\tau_Q$ for various system sizes and connectances, $c\in[0,1]$.
    Panels $(a)-(c)$: cumulant ratio  $\kappa_2(\rho_E)/\rho_E$ in the overdamped and underdamped cases for $N=401$ and for different values of $c$ in the KZ, intermediate and adiabatic regimes, respectively. In the KZ regime, the cumulant ratio is independent of $\tau_Q$ and increases universally with $c^2$. In the adiabatic regime, the ratio becomes proportional to the average decaying exponentially, as also highlighted with the additional fittings.
    Panels $(d)-(f)$: Second cumulant in the underdamped regime as a function of $\tau_Q$ in the same three regimes of $c$. Inside the KZ regime, the same power law is displayed as for the averages, but with a different universal rescaling with $r$. In the intermediate regime $c\in[0.15,0.5]$, a perfect scaling collapse is found as a function of $\tau_Q$.
    The adiabatic limit exhibits an exponential decay, matching approximately the square of the average. The structure of the legends is similar to that of Fig.~\ref{fig:averages_overdamped}.
    Panels $(g)-(i)$: Similar features with the corresponding power-laws in the overdamped regime with a reasonably faster adiabatic decay and less universal behavior for intermediate connectance values. An ensemble of $10^4$ trajectories is used.}
    \label{fig:kappa2_E}
\end{figure*}

Their numerical estimate is performed with the same number of $5\times 10^4$ trajectories. As shown in~Fig.~\figpanel{fig:cumulants_rho_E_n_2_n_1}{c,f} the $\kappa_3/\kappa_1$ and $\kappa_2/\kappa_1$ cumulant ratios of the one-dimensional defects follow precisely the same value as that of the $r=1$ case and of the TFIM. This agreement breaks down at slightly smaller values of the connectance, $c\approx0.1$, than the boundary of the universal KZ scaling of the averages. Correspondingly, $\kappa_2(n_1)$ and $\kappa_3(n_1)$ also become universal functions of the $r\tau_Q$ and $\sqrt r\tau_Q$ for the overdamped and underdamped regime, respectively (for further details see App.~\ref{app: second cumulants} and App.~\ref{app: third cumulants}).

Remarkably, in the fast quench regime, the proportionality is maintained with the same constant value as in the power law regime, as proposed in Ref. \cite{Xia2024}. This indicates that the shape of the distribution is the same under fast quenches and in the KZ regime.  However, this is not true at intermediate quenches as signaled by the non-universal peaks in a small interval of $\tau_Q$ before entering the KZ regime (see Apps.~\ref{app: second cumulants} and~\ref{app: third cumulants}).  
As demonstrated in Apps.~\ref{app: second cumulants} and~\ref{app: third cumulants}, in the adiabatic regime, the cumulants of $n_1$ and $n_2$ follow the same exponential law, given by the best numerical fit,
\begin{equation}\label{n_1_adiabatic_cums}
\begin{split}
    &\log\kappa_2(n_1)\sim\,-0.25r^{1/2}\tau_Q,\,\log\kappa_3(n_1)\sim\,-0.75r^{1/2}\tau_Q,\\      &\log\kappa_2(n_1)\sim\,1.4r^{1/3}\tau_Q,\,\log\kappa_3(n_1)\sim\,2.1r^{1/3}\tau_Q.
\end{split}
\end{equation}
This implies that up to numerical precision, the proportional nature of the cumulants turns into a power law relation up to exponential accuracy,
\begin{equation}\label{eq: n_1_cums}
    \kappa_2(n_1)\sim n^2_1,\quad \kappa_3(n_1)\sim n^3_1.
\end{equation}
This is in good agreement with the discussion of Sec.~\ref{sec: LMG} on the universal signatures in the intermediate regime. The exponential fits in Eq.~\eqref{n_1_adiabatic_cums} further demonstrate the universal rescaling $r^{\frac{z\nu}{1+z\nu}}\,\tau_Q$ in the adiabatic limit.

In the case of $\rho_E$, the cumulant ratios behave similarly to those of $n_1$, as observed in Fig.~\figpanel{fig:kappa2_E}{a} and in App.~\ref{app: third cumulants}. 
In the KZ regime, the ratio $\kappa_2(\rho_E)/\kappa_1(\rho_E) $ approximately follows constant plateaus, which get shifted as $ r$ is increased. This happens because, as for the averages, the excess energy depends on the actual rotor configuration and includes contributions from all rotors, including those at the sides of each kink.
Due to the graph theoretical normalization $1/Nr$, this effect boosts $\kappa_2(\rho_E)$ and $\kappa_3(\rho_E)$ with $r^2$ and $r^3$, respectively. This yields the following universal power laws,
\begin{align}\label{eq: E_third_cums}  &\kappa_2(\rho_E)\sim\left(r^{-3}\tau_Q\right)^{-1/4},\,\kappa_3(\rho_E)\sim\left(r^{-7}\tau_Q\right)^{-1/4},\\
&\kappa_2(\rho_E)\sim\left(r^{-5/2}\tau_Q\right)^{-1/3},\,\kappa_3(\rho_E)\sim\left(r^{-11/2}\tau_Q\right)^{-1/3}.
\end{align}


As shown in Fig.~\figpanel{fig:cumulants_rho_E_n_2_n_1}{a,b} and Fig.~\figpanel{fig:cumulants_rho_E_n_2_n_1}{d,e}, the corresponding $\kappa_3/\kappa_1$ and $\kappa_2/\kappa_1$ cumulant ratios acquire  scales $\sim r$ and $\sim r^2$, respectively leading to
\begin{align}\label{eq:E_cum_ratios}
    &\frac{\kappa_2(n_2)}{n_2}\sim\frac{\kappa_2(\rho_E)}{\rho_E}\sim\frac{\frac{n_1}{Nr^2}r^4}{rn_1}\propto c,\\
    &\frac{\kappa_3(n_2)}{n_2}\sim\frac{\kappa_3(\rho_E)}{\rho_E}\sim\frac{\frac{n_1}{N^2r^3}r^6}{rn_1}\propto c^2,
\end{align}
where the proportionality of the $n_1$ cumulants was explicitly written out. Note that the ratios exhibit precise scaling collapse as a function of $c$ in contrast to the universal $r$ dependence of the averages. Thus, energy cumulants naturally exhibit a crossover from the sub- to the super-Poissonan regime as $c$ is increased. 
This also implies that the corresponding distributions are no longer concentrated around the average and fluctuate asymmetrically.

In contrast to $n_1$, for fast quenches, the proportionality breaks down for the cumulant ratios of both $n_2$ and $\rho_E$, as displayed for $\kappa_2(\rho_E)$ in Fig.~\figpanel{fig:kappa2_E}{a,b,c} for the small $c\lesssim0.15$, intermediate $c\in[0.15,0.5]$, and large $c\gtrsim0.5$ connectance regimes, respectively.
In the adiabatic regimes, the same exponential dependence is found as for the averages, as demonstrated in Fig.~\figpanel{fig:kappa2_E}{b,c} for $\kappa_2(\rho_E)$, also highlighted by additional exponential fittings (see also App.~\ref{app: third cumulants} for $\kappa_3(\rho_E)$). However, the best numerical fits yield the following $r$ independent, universal exponential behavior,
 \begin{align}\label{eq: rho_E_cums_tau}
&\log\kappa_2(\rho_E)\sim\,1.6\tau_Q,\quad\log\kappa_3(\rho_E)\sim\,2.4\tau_Q,\\
&\log\kappa_2(\rho_E)\sim\,0.12\tau_Q,\quad\log\kappa_3(\rho_E)\sim\,0.18\tau_Q,
\end{align}
up to exponential accuracy in the overdamped and underdamped, respectively. Note that the $r$ independent behavior of the decay is dominated by small angel fluctuations rather than by the occurrence probability of the last pair of defects. 

By contrast, the cumulants of $n_2$ exhibit the same characteristics as those of $n_1$ up to exponential accuracy, as in this regime, defect densities are dominated by the vanishing occurrence probability of a single pair of defects.
Accordingly, the proportionality between the cumulants of $\rho_E$ and $n_2$ follow similar power law relations as in the case of $n_1$,
\begin{equation}\label{eq: powerlaw_cumratio}
\begin{aligned}
    &\kappa_2(\rho_E)\sim\rho^2_E~,&\kappa_3(\rho_E)\sim \rho^3_E,\\
    &\kappa_2(n_2)\sim n^2_2~,&\kappa_3(n_2)\sim n^3_2.
\end{aligned}
\end{equation}
Notably, keeping $r\sim N,\,c\lesssim0.1$ enforces the same power law relations as in Eq.~\eqref{eq: powerlaw_cumratio}.

Finally, let us also provide details about the surviving universal features in the intermediate regime.  As shown in Fig.~\figpanel{fig:kappa2_E}{a}, a perfect scaling collapse is found as a function of $\tau_Q$ in the underdamped regime. In the overdamped regime, universal signatures only survive up to not too large annealing times, $\tau_Q\lesssim10$, with the rescaling of $\sqrt r\,\tau_Q$, as shown in Fig.~\figpanel{fig:kappa2_E}{e}.

\section{\label{sec:conclusions}Conclusions}
We have investigated the annealing dynamics of finite-range interacting $\mathrm O(2)$ rotor networks on circulant regular graphs using numerical simulations and approximate analytical methods. This model provides an ideal testbed to benchmark quantum annealing devices and test the Kibble-Zurek mechanism in finite-range Ising models interacting with a thermal bath. Using the SVL model,  both the full-time evolution and equilibrium properties can be accessed. The latter is achieved by performing adiabatically slow time evolutions. 
In particular, we showed that for regular networks, the critical exponents and the average magnetization near the critical point remained invariant with respect to variations of the connectance. The only dependence was involved in the correlation length, which varied by a multiplicative factor.

Numerical results revealed that the KZ mechanism accurately described the universal power-law scaling of the density of point-like defects separating one-dimensional near-ferromagnetic domains. However, this measure could not account for the excess energy generated during the process. Therefore, the generalized graph defect density was introduced, which matched the excess energy density with high precision. For these quantities, the KZ power laws provided an accurate description in terms of the equilibrium critical exponents, however, with different regularity dependencies compared to the one-dimensional defects.  

The robustness of the universal KZ regime was explored as a function of the annealing time and the connectance. The defect and excess energy densities exhibited a transition to the universal fast quench regime below the corresponding regularity-dependent time scale. At the other extreme, adiabatic evolution dominated the dynamics above a system size- and regularity-dependent time scale. These threshold time scales identify a finite connectance for the universal breakdown of the KZ regime.
Above the adiabatic connectance scale, the defect and excess energy densities decayed exponentially. For both the one-dimensional and graph defects, similar universal properties were observed, albeit with regularity dependencies distinct from those in the KZ regime.  In the case of the excess energy, the adiabatic regime was dominated by small-angle deviations around the ground state. This resulted in a regularity-independent universal exponential decay in defect-free regimes of $\tau_Q$.

Finally, we analyzed the statistical distribution of the kink and energy densities by exploring the second and third cumulants. For the one-dimensional defect density, constant cumulant ratios were found in the KZ regime, with the same ratios as in the TFIM. Remarkably, this behavior persisted in the fast-quench regime as well. In the adiabatic regime, the ratios followed a power-law relation, which matched that of the graph defect density up to exponential accuracy.
In the case of the graph defect and excess energy densities, the proportionality between the cumulants acquired additional linear and quadratic regularity dependencies for the second and third cumulants, respectively. In contrast to the one-dimensional defects, this behavior broke down for fast quenches, while in the adiabatic regime, the cumulants followed the same power-law relation. Specifically, in this regime, the graph defect cumulants matched those of the one-dimensional defects up to exponential accuracy. However, in the case of the excess energy, different regularity-independent universal behavior was found. Similar to the averages, this behavior arose from the dynamics of small-angle deviations around the ferromagnetic directions in defect-free regimes of $\tau_Q$.

In short,  we have established the interplay of universality and its breakdown in the annealing dynamics of classical rotor networks with finite-range interactions. Our findings not only contribute to the understanding of nonequilibrium statistical mechanics of complex networks but also offer a benchmark for the performance of quantum simulators and annealing devices.   
As a step in this direction, during the writing of this manuscript, the preprint \cite{Cummins24VISA} reported the use of SVL on cyclic and random graphs as a benchmark in complex optimization by the Vector Ising Spin Annealer.

\acknowledgements
This research was funded by the Luxembourg National Research Fund (FNR), grant reference 17132054. 

\appendix

\section{Connectance dependence of the correlation length and time near the critical point}\label{app:corr_length_r}

In this appendix, we present an approximate analytical derivation of the spatial correlation function, unveiling its dependence on the connectance. Near the critical point, in the small angle limit, the Hamiltonian in Eq.~\eqref{eq:Hamiltonian} can be expanded as
\begin{equation}\label{eq:qudratic_Hamiltonian}
    H(t)\approx-\frac{J(t)}{2}\sum_{i,j=1}^NA_{ij}\theta_i\theta_j+h(t)\sum_{i=1}^N\theta^2_i.
\end{equation}

The correlation function is captured by directly considering the angle-angle equal-time two-point function, which is also expanded to leading order as
\begin{equation}
\begin{split}
G(d,t)&=\overline{\langle\sin\theta_n(t)\sin\theta_{n+d}(t)\rangle_\mathrm{traj}}\\
&\approx\overline{\langle\theta_n(t)\theta_{n+d}(t)\rangle_\mathrm{traj}}\\
&=\frac{1}{N}\sum_{n=1}^N\langle\theta_n(t)\theta_{n+d}(t)\rangle_\mathrm{traj},
\end{split}
\end{equation}
with $\overline{\cdots}$ denoting the spatial average over the rotors exploiting the spatial translational invariance of the system. The average over the random stochastic trajectories is replaced by the equilibrium canonical averaging with the  Boltzmann weight $e^{-\beta H(t)}$ as dictated by the fluctuation-dissipation theorem. Here, the inverse temperature $\beta=1/T=1000/J$ is given by the applied value of $T=0.001$ and where $k_B=1$ was set for convenience. To extract the equilibrium correlation length, the angles are expanded in the Fourier series as
\begin{equation}
    \theta_n=\frac{1}{\sqrt N}\sum_k\,e^{-2\pi ikn}\theta_k,\quad k=0,\,\frac{1}{N},\,\dots,\,1-\frac{1}{N}.
\end{equation}
This leads to the following Hamiltonian 
\begin{widetext}
\begin{equation}\label{eq: k_Hamiltonian}
\begin{split}
    H(t)&=
    -\frac{1}{2N}J(t)\sum_{k,k^\prime}\theta_k\theta_{k^\prime}\,\sum_{n=1}^N\,e^{2i\pi (k+k^\prime)n}\sum_{m=-r,\,m\neq 0}^r\,e^{-2i\pi  km}+h(t)\frac{1}{N}\sum_{n=1}^N\,e^{-2i\pi (k+k^\prime) n}\sum_{k,k^\prime}\theta_k\theta_{k^\prime}\\
    &=-J(t)\sum_k\,\theta_k\theta_{-k}\,\sum_{m=1}^r\cos\left(2\pi kn\right)+h(t)\sum_k\,\theta_k\theta_{-k}=\sum_k\,A_k\lvert\theta_k\rvert^2,
\end{split}
\end{equation}
\end{widetext}
where the identity $\frac{1}{N}\sum_{n=1}^N\,e^{-2i\pi kn}=\delta_{k,-k^\prime}$ was used. Furthermore, as the angles are real variables, the products were rewritten as $\theta_k\theta_{-k}=\lvert\theta_k\rvert^2$. Finally, we also introduced the overall constant of the quadratic Hamiltonian as
$A_k=\left[-2J\sum_{m=1}^r\cos\left(2\pi k m\right)+h\right]$. Here, for the sake of convenience, the time arguments were dropped everywhere.

Next, the Fourier decomposition is also applied to the correlation function 
\begin{equation}
    G(d)\approx\left\langle\frac{1}{N}\sum_{n=1}^N\theta_n\theta_{n+d}\right\rangle=\frac{1}{N}\sum_k\,e^{-2i\pi d k}\langle\lvert \theta_k\rvert^2\rangle,
\end{equation}
where the same steps were performed as above in Eq.~\eqref{eq: k_Hamiltonian} and the time dependence has been omitted for brevity. The single $k$ averages are simply evaluated by Gaussian integrations with variance $\beta A_k$
\begin{equation}\label{eq:Gaussian_integral}
    \langle\lvert\theta_k\rvert^2\rangle=\frac{1}{\beta A_k}=\frac{2TN}{h-2J\sum_{m=1}^r\cos\left(2\pi k m\right)}.
\end{equation}

For large enough values of $d$ capturing the long-distance fall-off of the correlations, the sum can be approximated by an integral with the new variable of $y=k$,
\begin{equation}
    G(d)\approx\frac{T}{\pi}\int_0^\infty\mathrm dy\frac{e^{iy\,d}}{h-2J\sum_{m=1}^r\cos\left(2\pi y m\right)}.
\end{equation}
Expanding to the leading order, the denominator of the integrand determines the decay of the correlations
\begin{widetext}
\begin{equation}\label{eq:G_l}
\begin{split}
    G(d)&\approx\frac{T}{\pi}\int_0^\infty\mathrm dy\frac{e^{iy\,d}}{h-2rJ+\frac{1}{12}r(r+1)(2r+1)y^2}
    \approx\frac{T}{2\pi}\int_0^\infty\mathrm dy\frac{e^{iy\,d}}{h-2rJ+\frac{r^3}{6}y^2}\\
    &\sim r^{-3}\int_0^\infty\mathrm dy\frac{e^{iy\,d}}{6\frac{h-2rJ}{r^3}+y^2}\propto r^{-3}\xi\,e^{-d/\xi}\propto r^{-2}e^{-d/\xi},
    \end{split}
\end{equation}
\end{widetext}
with the correlation length given by the pole of the denominator close to the critical point, where the order parameter  can be linearized, $\epsilon=h-2rJ\sim 2r\frac{t}{\tau_Q}$,
\begin{equation}
    \xi_\mathrm{eq}(r)\propto r^{3/2}\epsilon^{-1/2}.
\end{equation}
Note that the $r^{-2}$ constant factor emerged as the product of $r^{-3}$ in Eq.~\eqref{eq:G_l} and the overall $r$ dependence of the correlation length stemming from the regular part of the denominator, $\xi\sim r$.

As demonstrated in Fig.~\ref{fig:Corr_r} in the main text, near the critical point, $t_c\approx\tau/100$, the fitted exponential decay of the correlator increases linearly with the connectance, up to high precision, as observed in the insets for both the overdamped and underdamped regimes, verifying the above result. Intuitively, the above picture can also be understood as the correlation length being linearly scaled with the interaction range compared to the one-dimensional case. Thus, near the critical point, $\xi_\mathrm{eq}\sim r\left(t/\tau_Q\right)^{-1/2}\sim r^{3/2}\epsilon^{-1/2}$. Here, the $3/2$ power balances for the additional connectance dependence of the order parameter.

\section{Higher order corrections to criticality}
In this appendix, we analytically show that the higher order expansion of the Hamiltonian in Eq.~\eqref{eq:qudratic_Hamiltonian} close to the critical point does not affect the dynamical, $z$ and correlation length critical exponents, $\nu$. First, we show that the next-to-leading order correction does not modify the power-law divergence of the relaxation time-scale extracted from the Langevin equation, Eq.~\eqref{eq:overdamped}. For brevity, we restrict the exposition to the overdamped case, as all the steps can also be extended to the underdamped regime. The Langevin equation expanded up third order in the overdamped regime is given by
\begin{equation}
    \gamma\dot{\theta_i}+h(t)\theta_i - J(t)\sum_{\substack{j=-r \\ j\neq 0}} ^r\left(\theta_{i+j}-\theta_{i+j}\frac{\theta^2_i}{2}-\frac{\theta^3_{i+j}}{6}\right)+\xi_i(t)=0.
\end{equation}
Adapting the approximation of identical angles near the critical point, one obtains the equation for the scale of the relaxation time,
\begin{equation}
\label{eq:third_order_Langevin}
    \gamma\dot\theta+(h(t)-2rJ(t))\theta-\frac{2}{3}rJ(t)\theta^3=0.
\end{equation}
Starting from the leading order relation for the relaxation time, $\tau_0=(2rJ(t)-h(t))^{-1}$, Eq.~\eqref{eq:overdamped} and with the notation $\tau=\theta/\dot\theta$ for the new relaxation time stemming for the third order expansion, the correction is given by their difference, $\tau=\tau_0+\Delta\tau$. Writing back the $\tau_0$ to Eq.~\eqref{eq:third_order_Langevin}, one obtains
\begin{equation}
\begin{split}
    &\theta\left[(\tau_0+\Delta\tau)^{-1}-\tau^{-1}_0\right]-\frac{2}{3}rJ(t)\theta^3=0\\
    &\Rightarrow \theta^2\sim\frac{\Delta\tau}{\tau_0^2}.
    \end{split}
\end{equation}
This correction decays faster than $1/\tau_0$, as $\Delta\tau\ll\tau_0$ and $\Delta\tau\,\tau^{-2}_0\ll\tau^{-1}_0$,  thus not affecting in any way the critical properties of the relaxation time.

\begin{figure*}
\includegraphics[width=\linewidth]{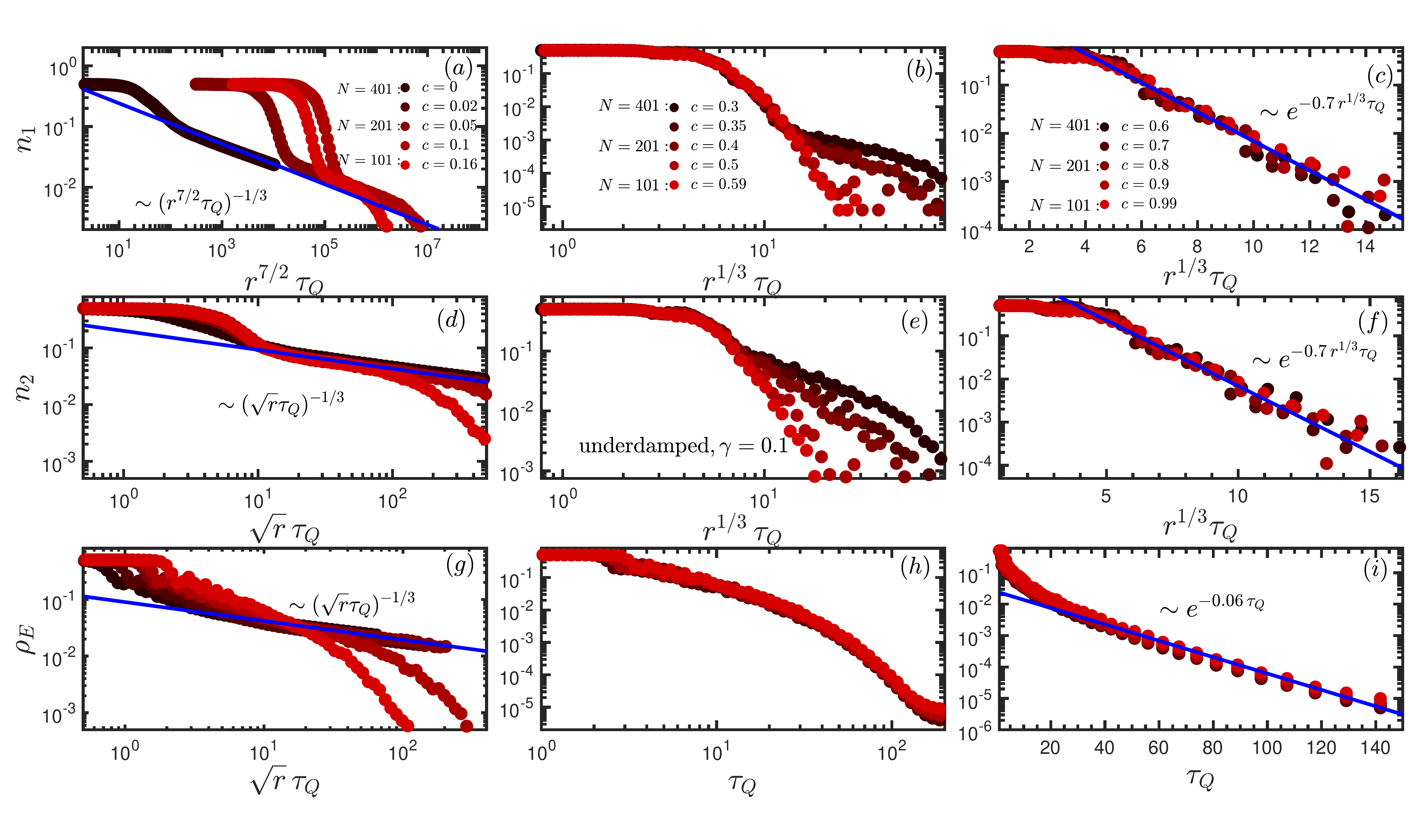}
\caption{Averages of defect and excess energy densities as a function of $\tau_Q$ in the overdamped regime with the connectance varying from the chain topology towards the fully connected limit for various system sizes. 
    Panels $(a)-(c)$: One-dimensional defect density for values $c\lesssim0.15$ The same power law is observed as in the chain topology with a universal regularity rescaling. For the intermediate and large connectances, a crossover is displayed towards the adiabatic regime, exhibiting a universal exponential shape. 
    Panels $(d)-(f)$: Similar observations for $n_2$ exhibiting the predicted universal KZ power laws of $r$ for $c\lesssim0.15$. Different universal rescaling emerges in the intermediate and adiabatic regimes. In the former, universal behavior breaks down for $\tau_Q\gtrsim10$. The structure of the legends is similar to that of Fig.~\ref{fig:averages_overdamped}.
    Panels $(g)-(i)$: Excess energy density, matching up to high precision the behavior of the finite range defect densities. In contrast to $n_2$, for $c\gtrsim0.15$, a perfect scaling collapse is observed as a function of $\tau_Q$. In all regimes,  curves involve averages over $10^3$ trajectories.\\
\label{fig:averages_underdamped}}
\end{figure*}

Next, the corrections to the correlation function are put to the test. The Hamiltonian in Eq.~\eqref{eq:qudratic_Hamiltonian} is expanded up to the fourth order as
\begin{equation}\label{eq:quartic_Hamiltonian}
\begin{split}
    H(t)&\approx-\frac{J(t)}{2}\sum_{i,j=1}^NA_{ij}\left(\theta_i\theta_j-\frac{1}{3}\theta^3_i\theta_j\right)\\
    &+h(t)\sum_{i=1}^N\theta^2_i-\frac{\theta^4_i}{12}.
    \end{split}
\end{equation}    
Note that the smallness of the angles is ensured by the low-temperature limit, $T\ll1$, as also obvious by the typical size of the angles set by its Fourier components in Eq.~\eqref{eq:Gaussian_integral}, $\langle\lvert\theta_k\rvert^2\rangle\sim T$.

\begin{figure*}
\includegraphics[width=\linewidth]{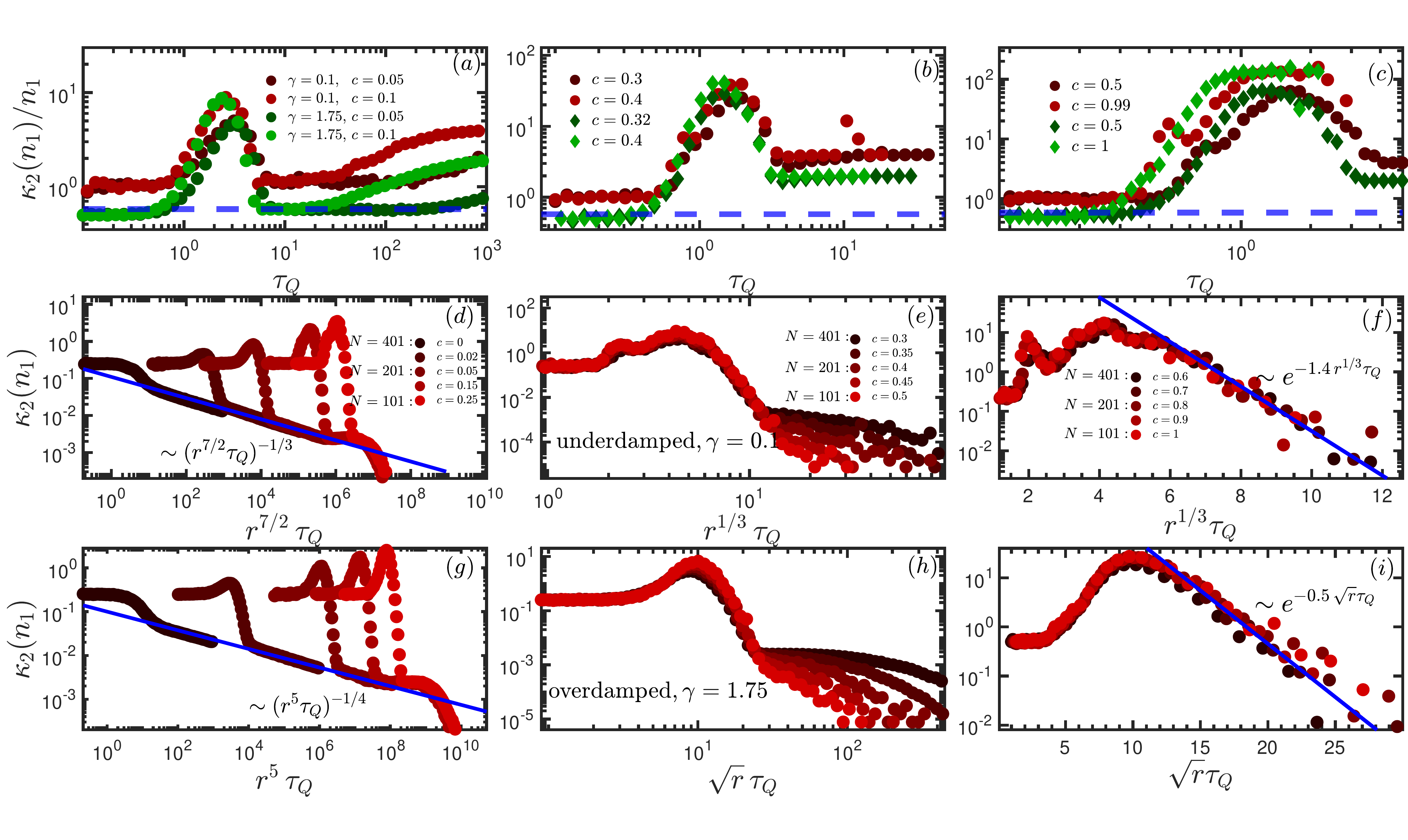}
\caption{Second cumulant of $n_1$ in the overdamped and underdamped regimes for the KZ, intermediate, and adiabatic connectance regimes, and for various system sizes. Panels $(a)-(c)$: Ratios of the first and second cumulants for the overdamped and underdamped regimes for the three connectance regimes, respectively, and for $N=401$. The ratio remains constant in the KZ and fast quench regimes. In the intermediate and adiabatic regimes, the proportionality breaks down, and an approximate power law relation is observed.
Panels $(d)-(f)$: Second cumulants of $n_1$ as a function of $\tau_Q$ in the underdamped regime $(d)$ showing the same power-law and crossover behavior in the KZ regime while exhibiting universal exponential decay in $(e)$ the intermediate and $(f)$ adiabatic ones. The universal behavior is governed by a different $r$ dependence and only survives up to $\tau_Q\lesssim10$ in the intermediate regime.
Panels $(g)-(i)$: Similar findings in the overdamped regime. Simulations involve an ensemble of $10^4$ trajectories. 
\label{fig:n_1_kappa_2}}
\end{figure*}
\begin{figure*}
\includegraphics[width=\linewidth]{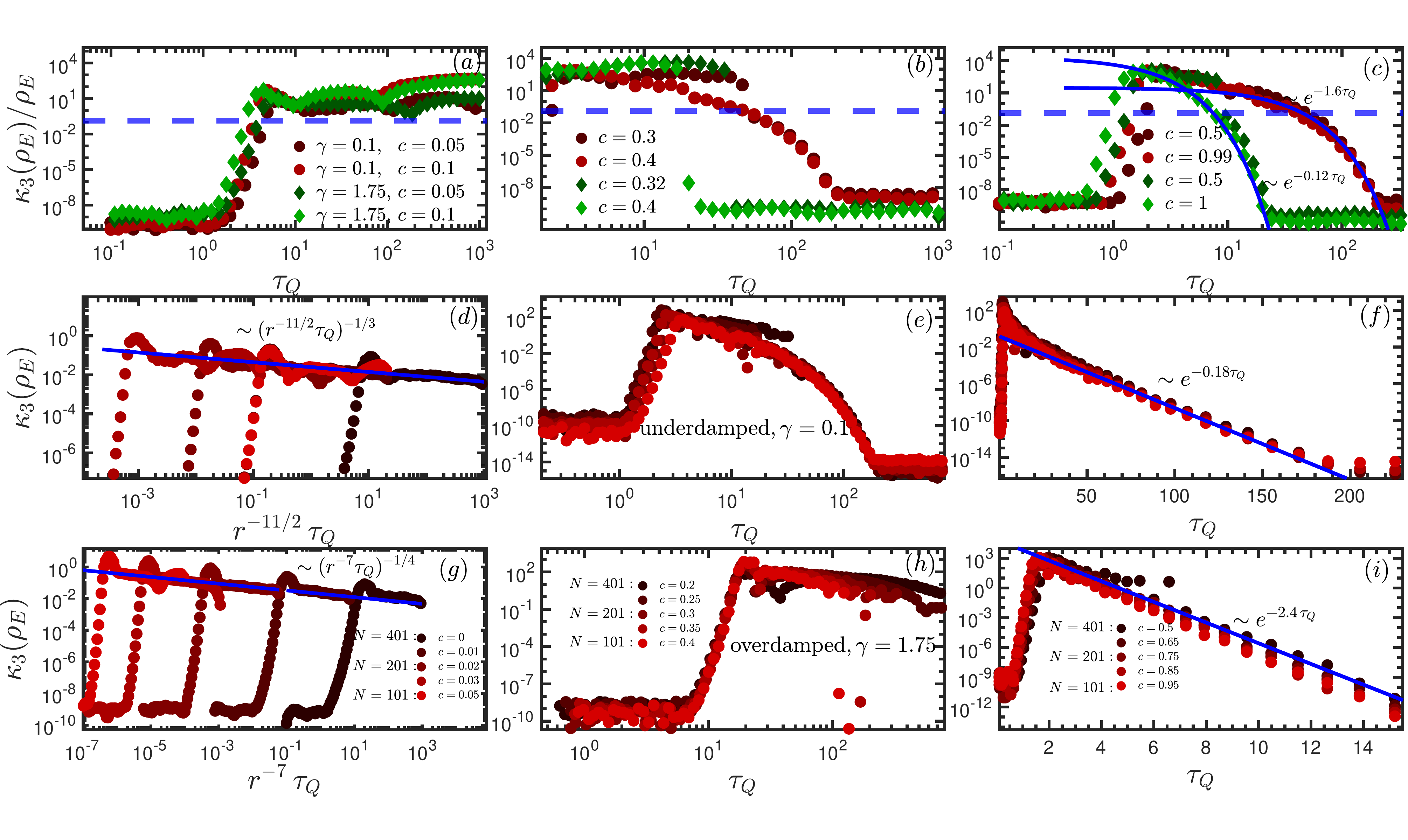}
\caption{Third cumulants of the excess energy density as a function of $\tau_Q$ for various system sizes, and connectances $c\in[0,1]$.
    Panels $(a)-(c)$: Cumulant ratios of $\kappa_3(\rho_E)/\rho_E$ both for the overdamped and underdamped cases, for different values of $c$ in the KZ, intermediate and adiabatic regimes, respectively and for $N=401$. In the KZ regime, the cumulant ratios are independent of $\tau_Q$ and increase universally with $c^2$. In the adiabatic regime, the ratio becomes proportional to the square of the average decaying exponentially, as also highlighted with the additional fittings.
    Panels $(d)-(f)$: Third cumulants in the underdamped regime as a function of $\tau_Q$ in the three connectance regimes. Inside the KZ regime, the same power law is displayed as for the averages, but with a different universal rescaling of $r$. Remarkably, in the $c\in[0.15,0.5]$ intermediate regime, perfect scaling collapse is found as a function of $\tau_Q$.
    The adiabatic limit exhibits an exponential decay, matching approximately the square of the average.
    Panels $(g)-(i)$: Similar features with the corresponding power-laws in the overdamped regime, exhibiting a faster adiabatic decay and less universal behavior for intermediate connectances. The reported data is extracted from an ensemble of $5\times 10^4$ trajectories.
\label{fig:E_kappa_3}}
\end{figure*}

In the correlator, the quartic corrections can only originate from the cubic expansion of the $\sin\theta_i$-s. However, the quartic expansion of the Hamiltonian will also enter the picture in terms of the normalization of the corresponding Boltzmann weights in the thermal average.
Starting with the latter, the normalization, i.e., the partition function in the quartic expansion is given by
\begin{widetext}
\begin{equation}
    Z\approx\int\prod_{k}\mathrm d\theta_k\,e^{-\beta \sum_kA_k\lvert\theta_k\rvert^2}\left(1+\frac{J}{6}\sum_{i,j=1}^NA_{i,j}\theta^3_i\theta_j-\frac{h}{12}\sum_{i=1}^N\theta^4_i\right)=\prod_kA^{-1}_k\left(1+\delta Z\right),
\end{equation}
\end{widetext}
where $\delta Z$ originates from the quartic correction. Here $A_k$ denotes the same function of $A_k=\left[-2J\sum_{m=1}^r\cos\left(2\pi k m\right)+h\right]$.
By decomposing the angle variable into Fourier series one can compute the $\delta Z$ with the same rules of Gaussian integrals as in Eq.~\eqref{eq:Gaussian_integral}
\begin{equation}
    \delta Z\approx\sum_{n=1}^N\sum_{k,q}\left[\frac{J}{6}\frac{1-e^{-d/\xi}}{e^{1/\xi}-1}\langle\lvert\theta_k\rvert^2\rangle\langle\lvert\theta_q\rvert^2\rangle-\frac{h}{12}\langle\lvert\theta_k\rvert^2\lvert\theta_q\rvert^2\rangle\right],
\end{equation}
which leads to
\begin{equation}
    \delta Z\sim r^{-4}T^2,
\end{equation}
where the factor $\frac{1-e^{-d/\xi}}{e^{1/\xi}-1}$ originates from the summation $\sum_{i,j=1}^N\,A_{ij}$ that translates to the summation of the exponentials inside the Fourier decomposition as $\sum_{r=1}^dº,e^{-r/\xi}=\frac{1-e^{-d/\xi}}{e^{1/\xi}-1}$.
In the last step, we wrote only the parametric dependence also highlighting the temperature acting as the small parameter.
The quartic correction of the $\sin\theta_i$ terms of the correlator takes the form of
\begin{equation}
\begin{split}
    G(d)\approx\frac{1}{N}\left[\sum_{n=1}^N\langle\theta_n\theta_{n+d}\rangle
    -\frac{1}{6}\langle\theta^3_n\theta_{n+d}\rangle\right]\left(1-\delta Z\right),
    \end{split}
\end{equation}
where the averaging is understood with the quadratic Hamiltonian, and the correction to the partition function has been involved in the leading order. Finally, we also compute the quartic term $\sum_n\langle\theta^3_n\theta_{n+d}\rangle$. For this, we employ again the Fourier decomposition of the angles to arrive at
\begin{equation}
    \sum_n\langle\theta^3_n\theta_{n+d}\rangle\sim\frac{1}{N}\sum_{k,q}\,e^{2\pi i kd}\langle\theta^2_k\rangle\langle\theta^2_q\rangle\sim T^2r^{-4}e^{-d/\xi},
\end{equation}
where in the last step we just used the result in Eq.~\eqref{eq:G_l}. The exponential decay is of the same order as in the quadratic approximation, but this term is also suppressed by the same factor $\sim r^{-4}T^2$ as the correction to the partition function.
Thus, both corrections only provide a subleading correction linear in the temperature as the quadratic is already proportional to $T$. Additionally, in the interesting $r\sim N$ regime the interaction range leads to a stronger suppression of the correction scaling as $\sim Tr^{-2}\sim TN^{-2}$
\begin{equation}
    G(d)\sim r^{-2}e^{-d/\xi}\left(1+O(Tr^{-2})\right).
\end{equation}

\section{Relaxation time in the generalized Langevin equation}\label{AppC}
In this appendix, we show that the coefficient of the relaxation time remains $r$ independent in a generalized Langevin dynamics, $\tau_0\sim r^0$. Following the strategy of the overdamped Langevin equation in Eq.~\eqref{eq:overdamped} we employ the leading order expansion and we assume that the $z\nu$ exponent originates from the generalized damping term,
    \begin{equation}
    m\partial^2_t{\theta_i}+\gamma\partial^q_t{\theta_i} +h(t)\theta_i + J(t)\sum_{\substack{j=-r \\ j\neq 0}} ^r\theta_{i+j}+\xi_i(t)=0.
\end{equation}
Here, we assume that the derivative, $q=1,2,\dots$ models the damping originating from the interaction with the environment.  Assuming that the typical angle values around the critical points are small and close to each other, one can write a relation similar to Eq.~\eqref{eq:overdamped},
\begin{equation}
    \tau\simeq\lvert \theta/\partial^q_t\theta\rvert^{1/q}\simeq \left\lvert\frac{\gamma}{h(t)-2rJ(t)}\right\rvert^{1/q}.
\end{equation}
 This leads to the relation $z\nu=1/q$. As a result, the finite-range extension of the interactions appears as a linear scaling factor only in the transverse field strength, and the relaxation time does not acquire any further explicit dependence on $r$, $\tau_0\sim r^0$.

\section{Average defect and excess energy density in the underdamped regime}\label{app: underdamped_averages}
In this appendix, we show the numerical results for the average defect density in the underdamped regime.
As shown in Fig.~\figpanel{fig:averages_underdamped}{a,d,g}, $n_1,\,n_2$ and $\rho_E$ follow precisely the predicted power laws as a function of $\tau_Q$ and $r$,~\eqref{eq: n_1_averages} in the KZ regime with $c\lesssim0.15$, 
\begin{align}    
    &n_1\propto\left( r^{7/2}\tau_Q\right)^{-1/3},\\
    &\rho_E\sim n_2\propto\left( r^{1/2}\tau_Q\right)^{-1/3}.
\end{align}
In the intermediate, $c\in[0.15,0.5]$ regime, universal signature only survive up to $\tau_Q\lesssim10$ as a function of $r^{1/3}\tau_Q$, as shown in Fig.~\figpanel{fig:averages_underdamped}{e,h}, respectively for $n_1$ and $n_2$. In the case of the excess energy, however, a remarkable scaling collapse is observed independently of $r$, as shown in Fig.~\figpanel{fig:averages_underdamped}{h}.
In the adiabatic limit, $c\gtrsim0.5$, $n_1$ and $n_2$ follow the same universal curve up to exponential accuracy as a function of $r^{1/3}\tau_Q$, while $\rho_E$ becomes independent of $r$ as governed by the dynamics of small angle deviations in defect-free regimes of $\tau_Q$,
\begin{align}        
    &n_1\sim n_2\sim e^{-0.7 r^{1/2}\tau_Q},\\
    &\rho_E\sim e^{-0.06 r^{1/4}\tau_Q}.
\end{align}

\begin{figure*}
\includegraphics[width=\linewidth]{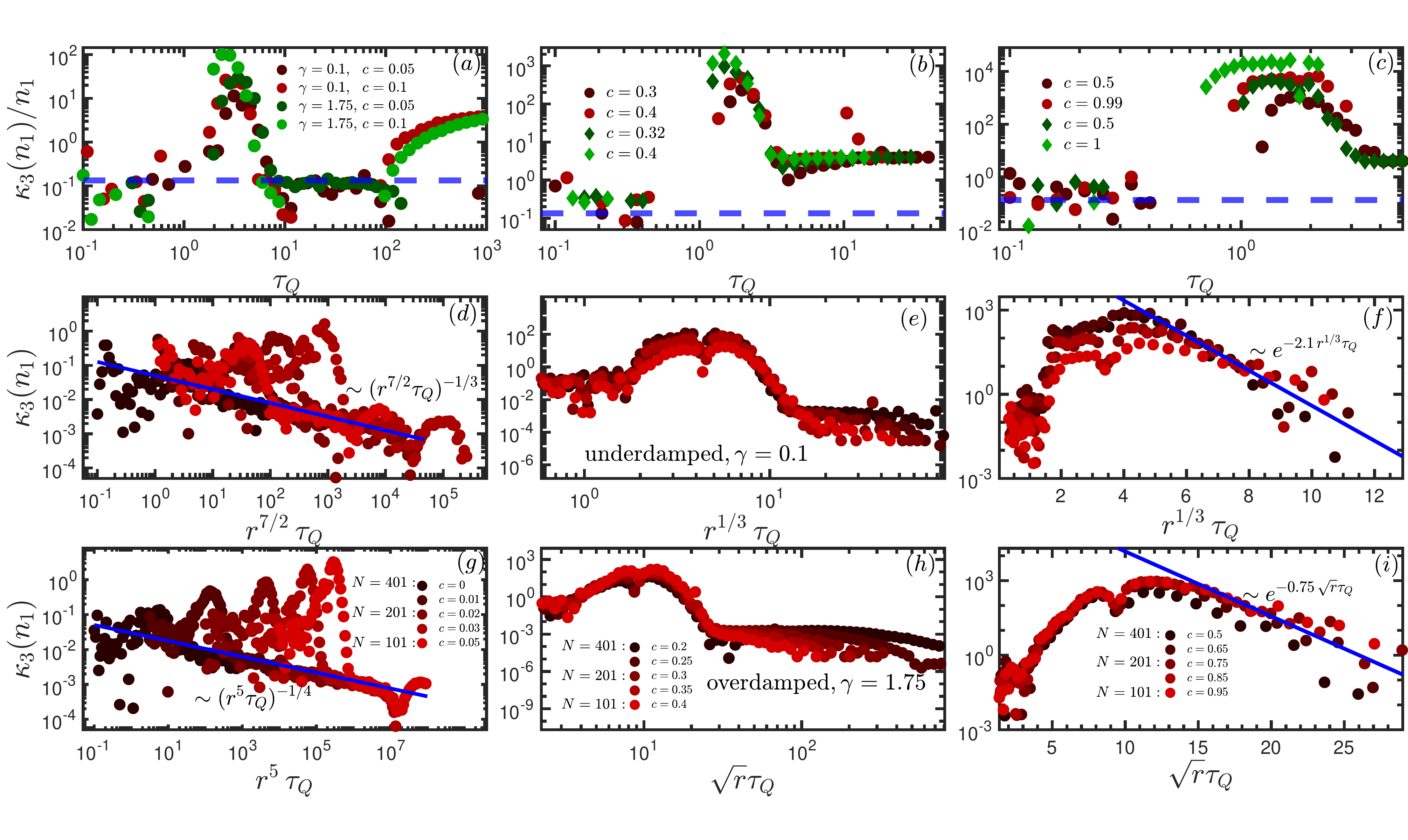}
\caption{Third cumulant of $n_1$ in the over- and underdamped cases for the KZ, intermediate, and adiabatic connectance regimes. Panels $(a)-(c)$: Ratios of $\kappa_3(n_1)/n_1$ for the over- and underdamped regimes, for $N=401$ and  the three connectance regimes, respectively. The ratio remains constant in the KZ and fast quench regimes. In the intermediate and adiabatic regimes, the proportionality breaks down, and an approximate power law relation is observed.
Panels $(d)-(f)$: Third cumulants of $n_1$ as a function of $\tau_Q$ in the underdamped regime showing the same power-law and crossover behavior in the KZ regime, $(d)$ while exhibiting universal exponential decay in the intermediate, $(e)$ and adiabatic ones, $(f)$. The universal behavior is governed by a different $r$ dependence and only survives up to $\tau_Q\lesssim10$ in the intermediate regime. 
Panels $(g)-(i)$: similar findings in the overdamped regime. The numerical results were obtained by averaging over $5\times 10^4$ trajectories.
\label{fig:n_1_kappa_3}}
\end{figure*}
\section{Second cumulant of the one-dimensional defects }\label{app: second cumulants}
In this appendix, a further numerical demonstration is provided for the $\tau_Q$ and $r$ dependence of the second cumulants of the one-dimensional defect density, $\kappa_2(n_1)$, and the corresponding ratio $\kappa_2(n_1)/n_1$. 

As demonstrated in Fig.~\figpanel{fig:n_1_kappa_2}{a} in the KZ regime, the proportionality is preserved independently of $r$ with the same value as in the TFIM, $\kappa_2(n_1)/n_1\approx0.578$. Remarkably, this universal behavior survives for fast quenches as well. Additionally, between the KZ and fast quench regimes, $r$-dependent non-universal peaks appear. In the intermediate ($c\in[0.15,0.5]$) and adiabatic ($c\gtrsim0.5$) regimes,  a sharp decrease is displayed after similar non-universal peaks, as shown in Fig.~\figpanel{fig:n_1_kappa_2}{b,c}. This is in good agreement with the power law relation between $\kappa_2(n_1)$ and $n_1$.

The dependence on $\tau_Q$ in the KZ regime is exhibited in Fig.~\figpanel{fig:n_1_kappa_2}{d,g} for the overdamped and underdamped regimes, respectively, following the universal power-laws predicted by the KZ mechanism in Eq.~\eqref{eq: n_1_averages}. In the intermediate regime, universal scaling collapse can only be achieved for $\tau_Q\lesssim10$ by the same regularity rescalings as in the adiabatic regime, $\tau_Q\rightarrow r^{\frac{z\nu}{z\nu+1}}\tau_Q$. In the latter case, curves with different values of $N$ and $r$ follow the same universal exponential decay. These characteristics are displayed in Fig. \figpanel{fig:n_1_kappa_2}{e,f} for the overdamped case and~in Fig. \figpanel{fig:n_1_kappa_2}{h,i} for the underdamped case.

\section{Third cumulants }\label{app: third cumulants}
In this appendix, the third cumulants are investigated as a function of $r,\,c$ and $\tau_Q$. As demonstrated in Fig.~\figpanel{fig:E_kappa_3}{a}, the excess energy exhibits the same features in the KZ regime as for $\kappa_2(\rho_E)$ in Sec.~\ref{sec:beyondKZM}. The cumulant ratios follow approximately constant lines that get shifted with $r$ beyond the fast quench regime. In this latter limit, $\kappa_3(\rho_E)/\rho_E$ converges to zero. As demonstrated in Fig.~\figpanel{fig:E_kappa_3}{b,c}, in the intermediate and adiabatic regimes of the connectance, the ratios exhibit an exponential decay, as also indicated by the additional fittings. This is in agreement with the power law relation between $\kappa_3(\rho_E)$ and $\rho_E$.

Furthermore, Fig.~\figpanel{fig:E_kappa_3}{d,g} shows the dependence on $\tau_Q$ following precisely the universal power laws in the KZ regime in Eq.~\eqref{eq: rho_E_cums_tau} for the overdamped and underdamped regimes, respectively. In the intermediate regime with $c\in[0.15,.5]$, universality breaks down around $\tau_Q\approx10$, while it survives in the underdamped regime, as shown in Fig.~\figpanel{fig:E_kappa_3}{e} and Fig.~\figpanel{fig:E_kappa_3}{h}, respectively. In the adiabatic regime with $c\gtrsim0.5$,  a universal, regularity-independent exponential decay is observed in agreement with the results on $\kappa_2(\rho_E)$ reported in Sec.~\ref{sec: LMG}.

In the case of $n_1$, similar features are found as in the case of $\kappa_2(n_1)$ in App.~\ref{app: second cumulants}. In the domain of the KZ power law scaling, the $\kappa_3(n_1)/n_1$ ratio follows precisely the same constant value as in the TFIM independently of $r$. Similar to the $\kappa_2(n_1)/n_1$ ratios, the proportionality also survives in the fast quench regime. A small interval of non-universal peaks also appears before the KZ regime, as shown in Fig.~\figpanel{fig:n_1_kappa_3}{a}.
In the intermediate and adiabatic regimes, the ratios exhibit exponential decay in agreement with the power-law relation, as shown in Fig.~\figpanel{fig:n_1_kappa_3}{b,c}.

As for the $\tau_Q$ dependence of $\kappa_3(n_1)$, the numerical results follow the same universal regularity and power law dependences in the KZ regime as for the average given in Eq.~\eqref{eq: n_1_averages}. These features are demonstrated in Fig.~\figpanel{fig:n_1_kappa_3}{d,g}. In the intermediate and adiabatic regimes, the universal exponential decay is displayed in agreement with Eq.~\eqref{eq: n_1_n_2_exp_decay}, as demonstrated in Fig.~\figpanel{fig:n_1_kappa_3}{e,f} and Fig.~\figpanel{fig:n_1_kappa_3}{h,i} for the overdamped and underdamped regimes, respectively.As in the case of $n_1$ and $\kappa_2(n_1)$, the universal scaling collapse in the intermediate regime as a function of $r^{\frac{z\nu}{z\nu+1}}\tau_Q$ only survives up to $\tau_Q\lesssim10$.

\newpage

\bibliography{SVLrotor}
\end{document}